\documentclass[10pt, a4paper]{article}
\usepackage{style}
\title{ Spin-orbital Jahn-Teller bipolarons }

\author[1,2]{Lorenzo Celiberti}
\author[1,3,4]{Dario Fiore Mosca}
\author[5]{Giuseppe Allodi}
\author[3,4]{Leonid V. Pourovskii}
\author[2]{Anna Tassetti}
\author[2]{Paola Caterina Forino}
\author[6]{Rong Cong}
\author[6]{Erick Garcia}
\author[7]{Phuong M. Tran}
\author[5]{Roberto De Renzi}
\author[7]{Patrick M. Woodward}
\author[6]{Vesna F. Mitrovi\'c}
\author[2]{Samuele Sanna}
\author[1,2,*]{Cesare Franchini}

\affil[1]{University of Vienna, Faculty of Physics and Center for Computational Materials Science, Vienna, Austria}

\affil[2]{Department of Physics and Astronomy 'Augusto Righi', Alma Mater Studiorum - Universit\`{a} di Bologna, Bologna, 40127 Italy}

\affil[3]{Centre de Physique Th\'eorique, Ecole polytechnique, CNRS, Institut Polytechnique de Paris, 91128 Palaiseau Cedex, France}

\affil[4]{Coll\`ege de France, Universit\'e PSL, 11 place Marcelin Berthelot,
75005 Paris, France }

\affil[5]{Department of Mathematical, Physical and Computer Sciences,
University of Parma, 43124 Parma, Italy}

\affil[6]{Department of Physics, Brown University, Providence, Rhode Island
02912, USA}

\affil[7]{Department of Chemistry and Biochemistry, The Ohio State University, Columbus, Ohio 43210, USA}

\affil[*]{e-mail: cesare.franchini@univie.ac.at}

\date{}
\begin{document}

\maketitle


\textbf{Polarons and spin-orbit (SO) coupling are distinct quantum effects that play a critical role in charge transport and spin-orbitronics.  
Polarons originate from strong electron-phonon interaction and are ubiquitous in polarizable materials featuring 
electron localization, in particular $\mathbf{3d}$ transition metal oxides (TMOs). On the other hand, the relativistic coupling between the spin and orbital angular momentum is notable in lattices with heavy atoms and develops in $\mathbf{5d}$ TMOs, where electrons are spatially delocalized. 
Here we combine ab initio calculations and magnetic measurements 
to show that these two seemingly mutually exclusive interactions are entangled 
in the electron-doped SO-coupled Mott insulator $\mathbf{Ba_2Na_{1-x}Ca_xOsO_6}$ ($\mathbf{0< x < 1}$), 
unveiling the formation of \emph{spin-orbital bipolarons}. 
Polaron charge trapping, favoured by the Jahn-Teller lattice activity, converts the Os $\mathbf{5d^1}$ spin-orbital $\mathbf{J_{eff}=3/2}$ levels,
characteristic of the parent compound $\mathbf{Ba_2NaOsO_6}$ (BNOO), into a bipolaron $\mathbf{5d^2}$ $\mathbf{J_{eff}=2}$ manifold, leading to the coexistence of different J-effective states in a single-phase material. 
The gradual increase of bipolarons with increasing doping creates robust in-gap states that 
prevents the transition to a metal phase even at ultrahigh doping, thus preserving the Mott gap across the entire doping range from $\mathbf{d^1}$ 
BNOO to $\mathbf{d^2}$ $\mathbf{Ba_2CaOsO_6}$ (BCOO).
}
\vfill

\begin{refsection}
The small polaron is a mobile quasiparticle composed of an excess carrier dressed by a phonon cloud~\cite{franchini_polarons_2021,Alexandrov2010book,landau_uber_1933, emin_2012}. It is manifested by local structural deformations and flat bands near the Fermi level and is significant for many applications including
 photovoltaics~\cite{miyata_large_2017, guzelturk_visualization_2021,PhysRevLett.110.196403}, 
rechargeable ion batteries~\cite{luong_small_2022}, surface reactivity~\cite{DiValentin2009, PhysRevLett.122.016805,Pastor2022},
high-$\mathrm{T_c}$ superconductivity~\cite{zhao_evidence_1997} and colossal magnetoresistance~\cite{lee_polaron_1997}.
Coupling polarons with other degrees of freedom can generate new composite quasiparticles, such as magnetic~\cite{Teresa1997}, Jahn-Teller (JT)~\cite{Hock,allodi_ultraslow_2001}, ferroelectric~\cite{Miyata2018} and 2D polarons~\cite{Sio2023}, to name just a few.
The main driving forces favoring polaron formation: phonon-active lattice, electronic correlation, and electron-phonon coupling, are realized in 3d TMOs, which represent a rich playground for polaron physics~\cite{franchini_polarons_2021, Stoneham2007,PhysRevB.99.235139}.
In 5d TMOs, instead, charge trapping is hindered by the large d-bandwidth and associated weak electronic correlation, making polaron formation in a 5d orbital an unlikely event~\cite{Reticcioli2022-lf}. 

The recent discovery of SO coupled Mott insulators~\cite{Jackeli2009}, where the gap is opened by the cooperative action of strong SO coupling and appreciable electronic correlation, has paved the way for the disclosure of novel, exciting quantum states of matter~\cite{Witczak-Krempa2014,Rau2016}.
The coexistence of SO coupling and electronic correlation in the same TMO raises the possibility of conceptualizing a SO polaron~\cite{xing_localized_2020, arai_multipole_2022}, provided that the correlated relativistic background develops in a structurally flexible lattice.
These conditions are met in the double perovskite 
$\mathrm{Ba_2NaOsO_6}$ (BNOO)~\cite{lu_magnetism_2017}. With a SO coupling strength $\lambda$ of 0.3~eV, a large on-site Hubbard $U$ of 3.4~eV and sizable JT vibration modes~\cite{Erickson2007,PhysRevB.97.224103, cong_first_2020, fiore_mosca_interplay_2021}, BNOO represents the ideal candidate for questing 5d \emph{spin-orbital polarons}. 

\begin{figure}
    \centering
    \begin{subfigure}{\linewidth}
        \caption{}
        \label{fig:polaron}
        \centering
        \includegraphics[width=180mm]{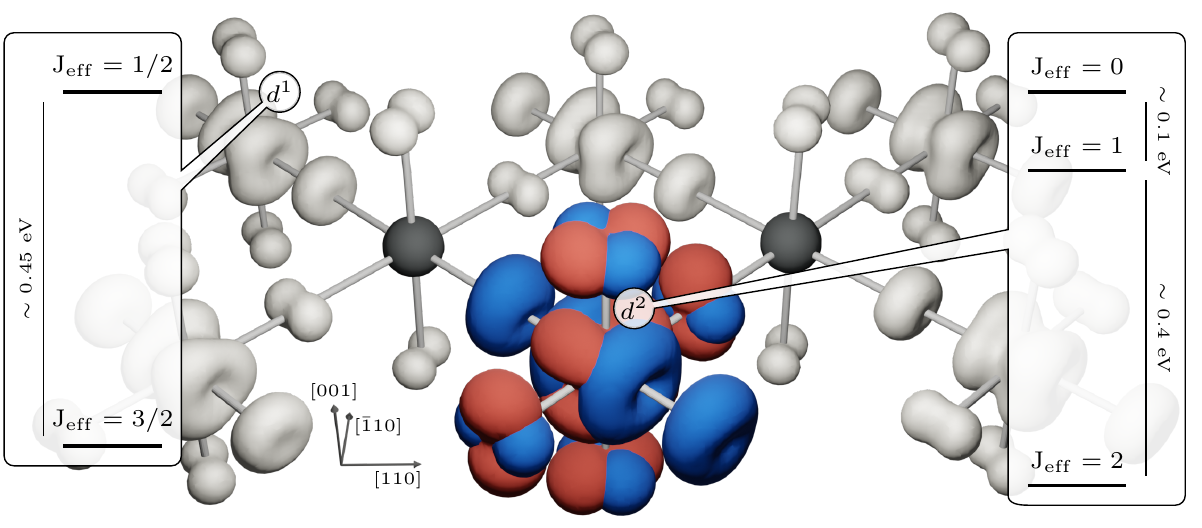}
    \end{subfigure}
    \vfill
    \begin{subfigure}{\linewidth}
        \caption{}
        \label{fig:dosbstruct}
        \centering
        \includegraphics[width=\linewidth]{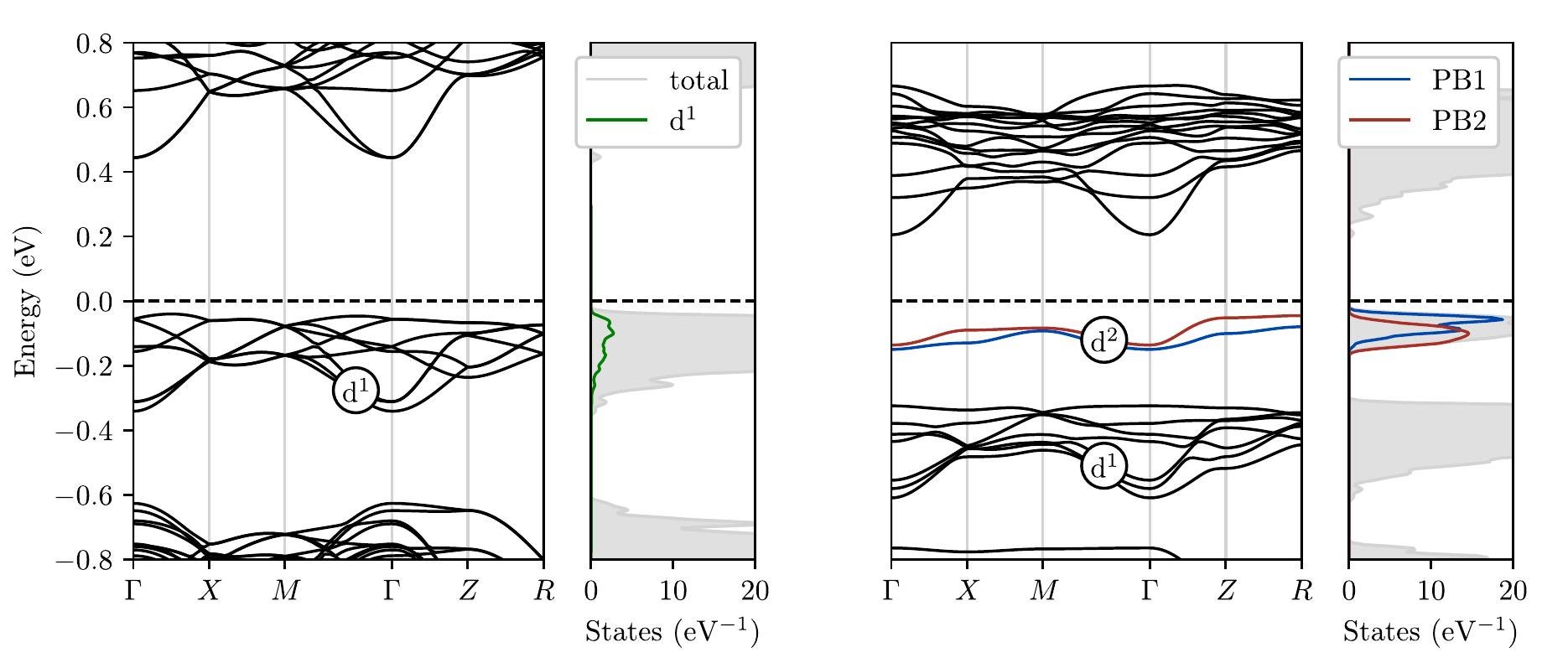}
    \end{subfigure}
    \caption{{\textbf{Spin-orbit $\mathbf{d^2}$ bipolaron in $\mathbf{Ba_2Na_{0.875}Ca_{0.125}OsO_6}$}}.
    \textbf{(a)} DFT charge density isosurface of the occupied Os $t_{2g}$ bands, showing the formation of $\mathrm{d^2}$ a J$_{\mathrm{eff}}=2$
    bipolaron coexisting with $\mathrm{d^1}$ J$_{\mathrm{eff}}=\frac{3}{2}$ sites characteristic of pristine $x=0$ BNOO. Blue and red lobes refers to the entangled bipolaronic PB1 and PB2 bands displayed in panel b.
    The J$_{\mathrm{eff}}$ spin-orbital levels are obtained from DFT+HI. \textbf{(b)} Band structure and relative density of states of pristine BNOO ($x=0$, left) characterized by a multiband manifold of $d^1$ states, and bipolaronic $\mathrm{Ba_2Na_{0.875}Ca_{0.125}OsO_6}$ (right) with a localized $\mathrm{d^2}$ bipolaronic level below the Fermi level composed by two entangled $\mathrm{d^1}$ bands (PB1 and PB2).}
    \label{fig:elProp}
\end{figure}

BNOO is a Mott insulator with a low temperature 
canted antiferromagnetic (cAFM) ordered phase below 7~K, 
where SO splits the effective $l=1$ $t_{2g}$ levels on the Os$^{7+}$ $\mathrm{d^1}$ ion into a lower $\mathrm{J_{eff}}=3/2$ ground state and a doublet $\mathrm{J_{eff}}=1/2$~\cite{lu_magnetism_2017, fiore_mosca_interplay_2021, iwahara_spin-orbital-lattice_2018} (see Fig.~\ref{fig:elProp}(a)). Injecting electrons in $\mathrm{Ba_2Na_{1-x}Ca_{x}OsO_6}$ by chemical substitution of monovalent Na with divalent Ca ions does not cause the collapse of the Mott gap, which remains open up to full doping, when all $\mathrm{d^1}$ sites are converted in $\mathrm{d^2}$~\cite{kesavan_doping_2020}. 
This indicates that excess charge carriers do not spread uniformly in the crystal, forming a metallic state, but rather should follow a different fate. Here we provide evidence that the addition of excess electrons at a local $\mathrm{J_{eff}}=3/2$ site produces the formation of SO/JT entangled $\mathrm{J_{eff}}={2}$ bipolarons, which block the onset of a metallic phase. 

To gain insights on the effect of electron doping in BNOO we have performed Density Functional Theory (DFT) calculations on a $\mathrm{Ba}_2\mathrm{Na_{1-x}Ca_xOsO}_6$ supercell containing eight Os atoms at $x=0.125$, corresponding to one extra electron per supercell.
Fig.~\ref{fig:elProp}(a) shows that the extra charge is trapped at a $\mathrm{d^1}$ Os site, leading to a local modification of the electronic configuration from $\mathrm{d^1 \to d^2}$. The surrounding OsO$_6$ oxygen octahedron expands isotropically by a few $10^{-2}$~\AA, and new nearly flat bands develop in the mid gap region, all hallmarks of small polaron formation. This is confirmed by the local polaronic charge displayed in Fig.~\ref{fig:elProp}(a). 
The delocalized alternative metal phase, with the excess charge equally distributed over all Os sites~\cite{reticcioli_small_2019}, is less stable than the small polaron phase by 134~meV.

Electron trapping in TMOs generally occurs in an empty $\mathrm{d^0}$ manifold at the bottom of the conduction band, causing a $\mathrm{d^0 \to d^1}$ transition at the trapping site associated with one mid-gap flat band. In BNOO, where the $\mathrm{5d}$ orbital is singly occupied and strongly hybridized with Oxygen $\mathrm{p}$ states~\cite{kesavan_doping_2020}, we observe a conceptually different mechanism. 
In the undoped phase the fully occupied $\mathrm{d^1}$ states are grouped among the topmost valence bands 
(see Fig.~\ref{fig:elProp}(b)) and each site contributes equally to the density of state (DOS) (green line, bandwidth $W\approx 0.3$~eV). The chemically injected excess electron goes to occupy an empty d band at the bottom of the conduction manifold, which is shifted into the gap
and couples with the original $\mathrm{d^1}$ band at the same lattice site, forming a local $\mathrm{d^2}$ configuration (PB1 and PB2 bands in Fig.~\ref{fig:elProp}(b)) well separated by the remaining $\mathrm{d^1}$ bands. 
The resulting $\mathrm{d^2}$ dual-polaron complex can be assimilated to a bipolaron~\cite{Alexandrov1996}, as evident from the charge isosurface shown in Fig.~\ref{fig:elProp}(a), where the PB1 and PB2 orbitals are interwoven together.

\begin{figure}

    \centering
    \includegraphics[width=\linewidth]{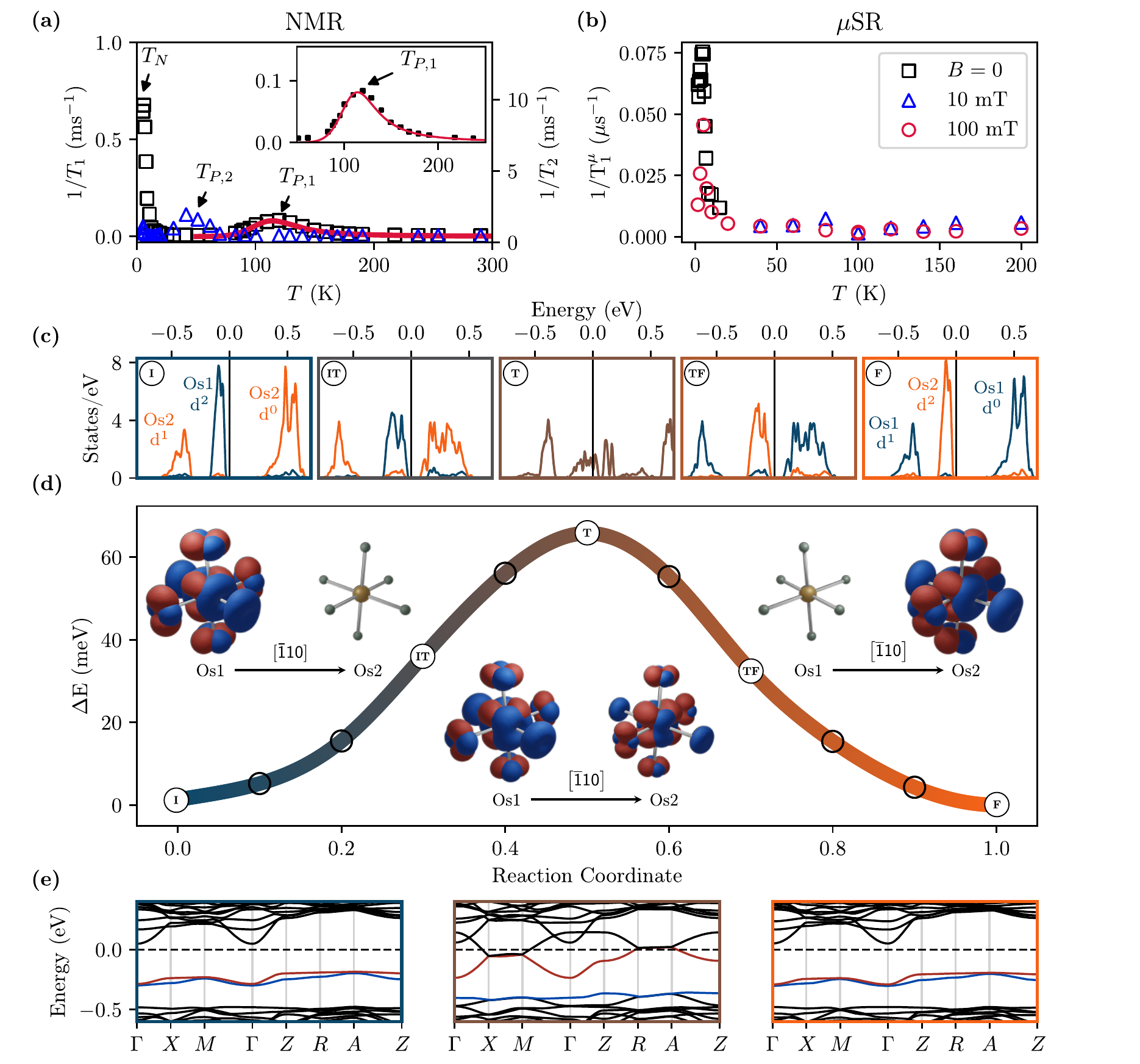}
    \caption{{\textbf{Polaron hopping: experiment and DFT}} \textbf{(a)} NMR spin-lattice (square) and spin-spin (triangles) relaxation rates showing an anomalous peak at 130~K and 50~K due to a dynamical process; the curve fit (solid line) to a thermally activated BPP model provides an activation energy $E_a=74(2)$~meV. 
    \textbf{(b)} $\mu$SR data showing only a peak due to the magnetic transition but no high temperature anomalous feature corresponding to the NMR one. 
    \textbf{(c)} Evolution of the density of states around the Fermi level for selected snapshots across the hopping path displayed in panel d, projected onto the initial (I, dark blue) and final (F, orange) Os sites. The five plots correspond to reaction coordinate equal to 0.1, 0.3, 0.5, 0.7 and 0.9. The $\mathrm{d^2}$ bipolaron charge is gradually transferred from the initial and final hosting sites. At the transition state (T, at 0.5) the charge is equally distributed between both Os sites giving rise to an adiabatic weakly metallic transient state (brown). 
    \textbf{(d)} Potential energy for a bipolaron migrating from I to F with the charge density projected on the two neighboring Os atoms, using a color gradient from blue (bipolaron fully localized in I) to orange (bipolaron fully localized in F). The insets show the charge density isosurface decomposed over the bipolaron bands PB1 (blue) and PB2 (red). The resulting hopping barrier, 0.66~meV, is in excellent agreement with the experimentally-derived activation energy.
    \textbf{(e)} band structure around the Fermi level at the initial (I), transition (T) and final (F) point of the hopping process.}
    \label{fig:hopping}
\end{figure}

Polaron formation is confirmed by
$^{23}$Na nuclear magnetic resonance (NMR) and muon spin rotation ($\mu$SR) measurements on a BNOO sample having $12.5\%$ $\mathrm{Ca}$ concentration shown in Fig.~\ref{fig:hopping}(a) and (b), respectively.
NMR shows an anomalous peak at $T_{P,1}\approx 130$~K in the spin-lattice relaxation rate $1/T_1$ (squares), well above the temperature associated to the magnetic transition (6.8~K); correspondingly, a peak is observed in the spin-spin relaxation rate $1/T_2$ at $T_{P,2}\approx 50$~K (triangles). 
Since the fast paramagnetic fluctuations are beyond the frequency window employed and no specific magnetic interaction is expected in the explored regime~\cite{garcia2022}, we attribute the NMR anomalous peaks to a charge-related thermally activated process, such as that associated with the small-polaron dynamics. This dynamical process drives electric field gradient (EFG) fluctuations which are probed by the quadrupolar interaction with the $^{23}$Na nuclear quadrupole. A peak is expected in $1/T_1$ when the frequency of the EFG fluctuations $\nu=1/\tau_c$ (being $\tau_c$ the fluctuation correlation time) matches the Larmor frequency (here $\omega_0\approx 5 \cdot 10^8$ s$^{-1}$), while a peak in $1/T_2$ is anticipated when $\tau_c$ is of the order of the experimental NMR echo delay time (here of the order of microseconds). In order to confirm that the origin of the observed peaks in the NMR rates can solely be associated with   the small-polaron dynamics, we have performed the $\mu$SR measurements, which are only sensitive to magnetic fluctuations.   
The $\mu$SR results exhibit strong critical relaxation rates at the magnetic transition temperature, 7 K, but no further relaxation peak above, in agreement with the quadrupolar polaronic mechanism~\cite{allodi_ultraslow_2001}, since the spin 1/2 muon is not coupled to EFGs.
The  anomalous peak in NMR $1/T_1$  temperature dependence is fitted using a Bloembergen-Purcell-Pound-like (BPP) model for quadrupolar spin-lattice relaxation~\cite{bloembergen_relaxation_1948, andrew_spin-lattice_1961}
\begin{equation}
    \frac{1}{T_1} = \Delta^2 \left[ \frac{\tau_c}{1+(\omega_0\tau_c)^2} + \frac{4\tau_c}{1+(2\omega_0\tau_c)^2} \right],
    \label{eq:bppmodel}
\end{equation}
where $\Delta^2$ is the second moment of the perturbing quadrupole-phonon coupling and the correlation time $\tau_c=\tau_0\exp(E_a/kT)$ is expressed in terms of the activation energy $E_a$ and the characteristic correlation time $\tau_0$ of the dynamical process. 
The resulting fitting curve is represented by the solid line in Fig.~\ref{fig:hopping}(a) and predicts a dynamical process with activation energy of $E_a=74(2)$~meV, $\tau_0=0.7(0.2)$~ps and $\Delta^2 = 2.75(6)\times 10^{10}$~s$^{-2}$.  

The energy barrier extracted from $1/T_1$ is in good agreement with the activation energy predicted by DFT for a thermally activated adiabatic hopping, $66$~meV, estimated at the same doping level in the framework of the Marcus-Emin-Holstein-Austin-Mott (MEHAM) theory~\cite{marcus_electron_1993, emin_studies_1969}.
The energy path between two energetically equivalent initial (I) and final (F) polaron sites Os1 and Os2 along the [$\bar{1}$10] direction, constructed with a linear interpolation scheme (LIS)~\cite{deskins_electron_2007}, is shown in Fig.~\ref{fig:hopping}(d). The hopping is a complex mechanism involving a three electron process:
at the initial stage (I) the DOS (see Fig.~\ref{fig:hopping}(c)) is characterized by the $\mathrm{d^2}$ polaron peak at the Os1 site (blue) and the unperturbed $\mathrm{d^1}$ and $\mathrm{d^0}$ bands at the final Os2 site (orange lines);
When the hopping process starts, the Os1-$\mathrm{d^2}$ and Os2-$\mathrm{d^0}$ bands get progressively closer, and a fraction of the polaronic charge in Os1 transfers to the empty band in Os2. At the transition (T) state the polaron charge is equally distributed between Os1 and Os2 resulting in a local weakly metallic state (see Fig.~\ref{fig:hopping}(c) and (e)), as expected from an adiabatic hopping process~\cite{deskins_electron_2007}. At this point a reverse mechanism begins: the original Os2-$\mathrm{d^1}$ and (now filled) Os2-$\mathrm{d^0}$ merges to form a $\mathrm{d^2}$ polaron in Os2, whereas the original Os1-$\mathrm{d^2}$ is depleted by one electron and generates a $\mathrm{d^1}$ band below the polaron peak. As a result, the DOS at the final point F is symmetrical to I (compare panels I and F in Fig.~\ref{fig:hopping}(c) and corresponding polaron charge isosurfaces in the insets of Fig.~\ref{fig:hopping}(d)).

Next we investigate the nature of the polaron, unravelling an intermingled action of SO and JT-distortions in determining the energy levels and degree of stability of the polaron~\cite{streltsov_jahn-teller_2020}. 
As single-particle approach, DFT is bound to predict $jj$-coupled levels~\cite{khomskii_orbital_2021}, where the total angular momentum J is the vector addition of the single-electron angular momentum $j$. Indeed, the $\mathrm{d^2}$ polaron occupation matrix computed by projecting the Kohn-Sham energy levels onto the $\mathrm{d^2}$ polaronic subspace using spinorial projected localised orbitals~\cite{mosca_mott_2023} shows that the trapped electrons occupy two single-particle $\mathrm{J_{eff}}=3/2$ levels (see Tab.~\ref{tabsm:occmat} and Fig.~\ref{figsm:colorocc} in SI) corresponding to the PB1 and PB2 bands reported in Fig.~\ref{fig:elProp}(b).
To compute the $\mathrm{d^2}$ two-electron levels we employed Dynamical Mean-Field Theory (DMFT) within the Hubbard-I (HI) approximation 
applied to the DFT lattice structure relaxed with the polaronic site. 
This many-body approach finds that the two electrons forming the $\mathrm{d^2}$ polaron occupy $\mathrm{J_{eff}}=2$ LS-coupled levels (see Fig.~\ref{figsm:2ellev} and Fig.~\ref{figsm:hidos}), separated from the excited $\mathrm{J_{eff}}=1$ triplet by a SO gap of about $0.4$~eV, as schematized in Fig.~\ref{fig:elProp}(a). The non-polaronic $\mathrm{d^1}$ sites (gray isosurfaces in Fig.~\ref{fig:elProp}(a)) preserves a $\mathrm{J_{eff}}=3/2$ ground state, as in the pristine material~\cite{fiore_mosca_interplay_2021}.
Regardless the specific type of coupling, $jj$ or LS, both
DFT and DMFT predict spin-orbital $\mathrm{J_{eff}}$ states, clearly indicating that the polaron is integrated into the SO-Mott background, it exhibits an individual spin-orbital state, and does not break the preexisting $\mathrm{J_{eff}}=3/2$ state at the other Os sites. This leads to the coexistence of two distinct SO-Mott quantum states in the same material, a hitherto unreported physical scenario.

\begin{figure}[ht]
    \centering
    \begin{subfigure}{\linewidth}
       \centering
       \includegraphics[width=\linewidth]{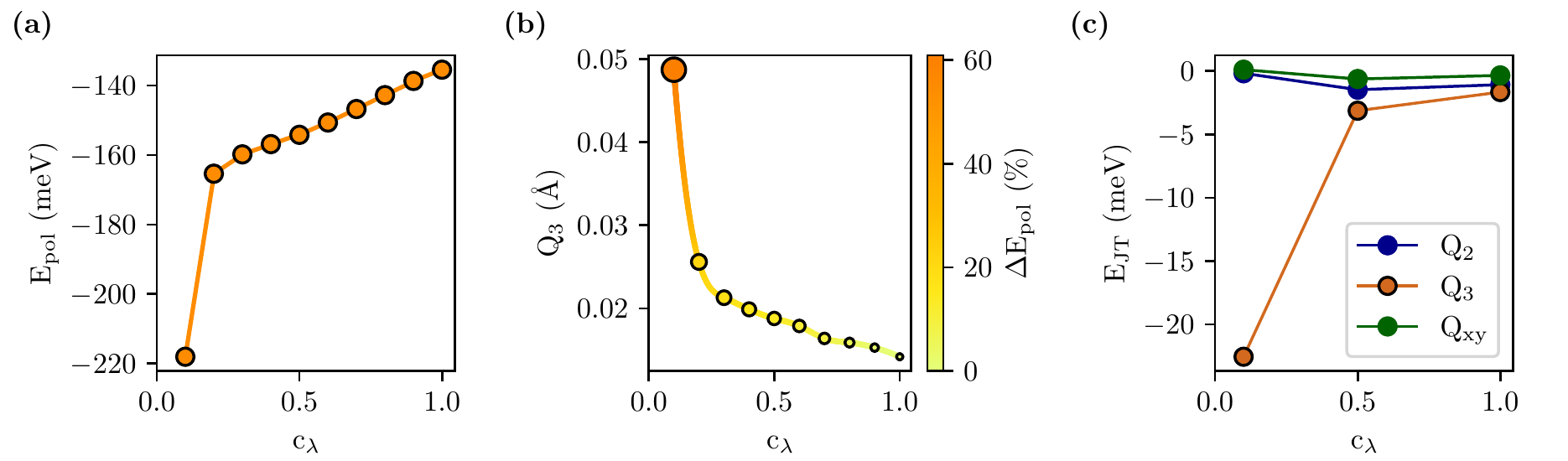}
    \end{subfigure}
    \vfill
    \addtocounter{subfigure}{3}
    \begin{subfigure}{0.3\linewidth}
    \centering
    \caption{}
    \includegraphics[width=0.8\linewidth]{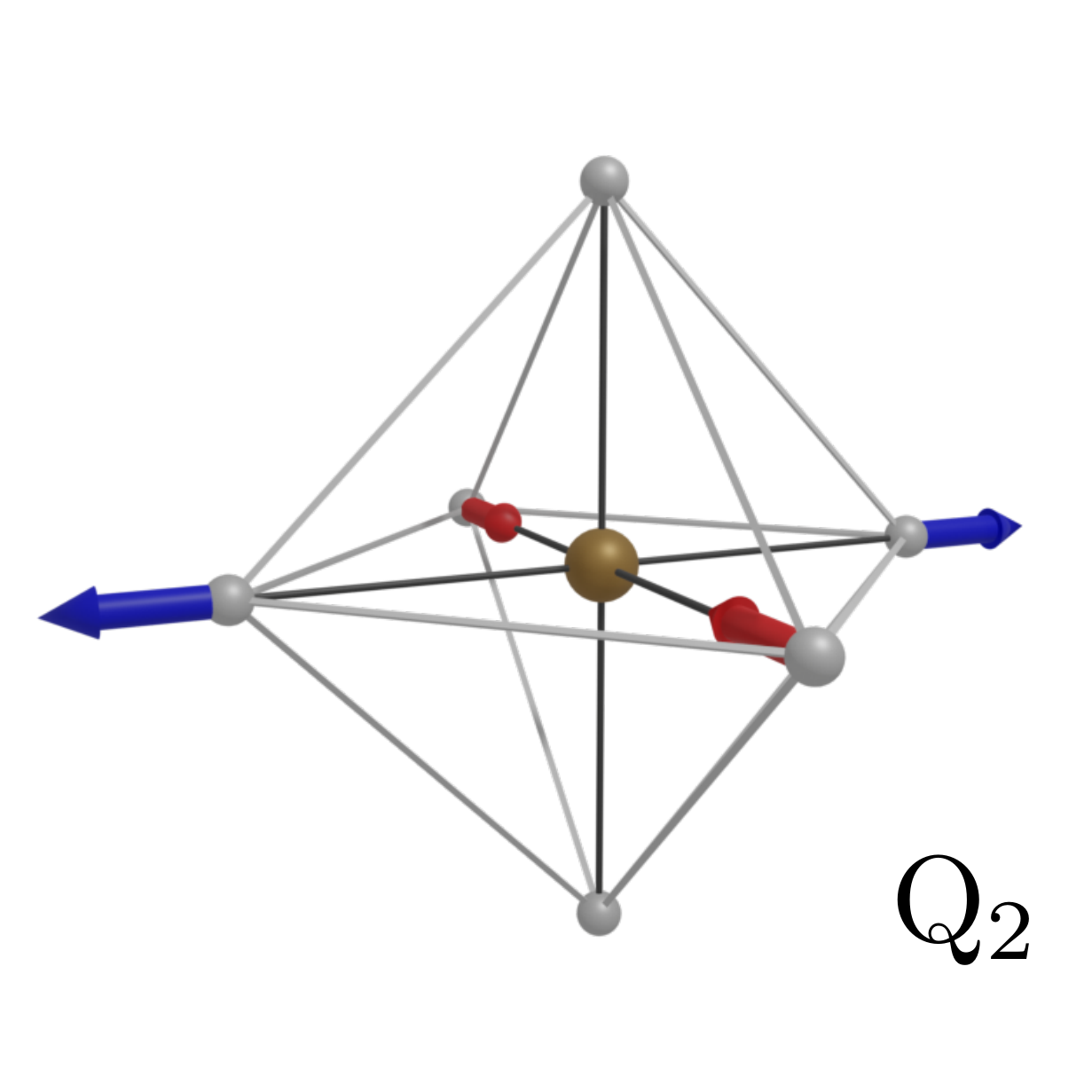}
    \end{subfigure}
    \hfill
    \begin{subfigure}{0.3\linewidth}
    \caption{}
    \centering
    \includegraphics[width=0.8\linewidth]{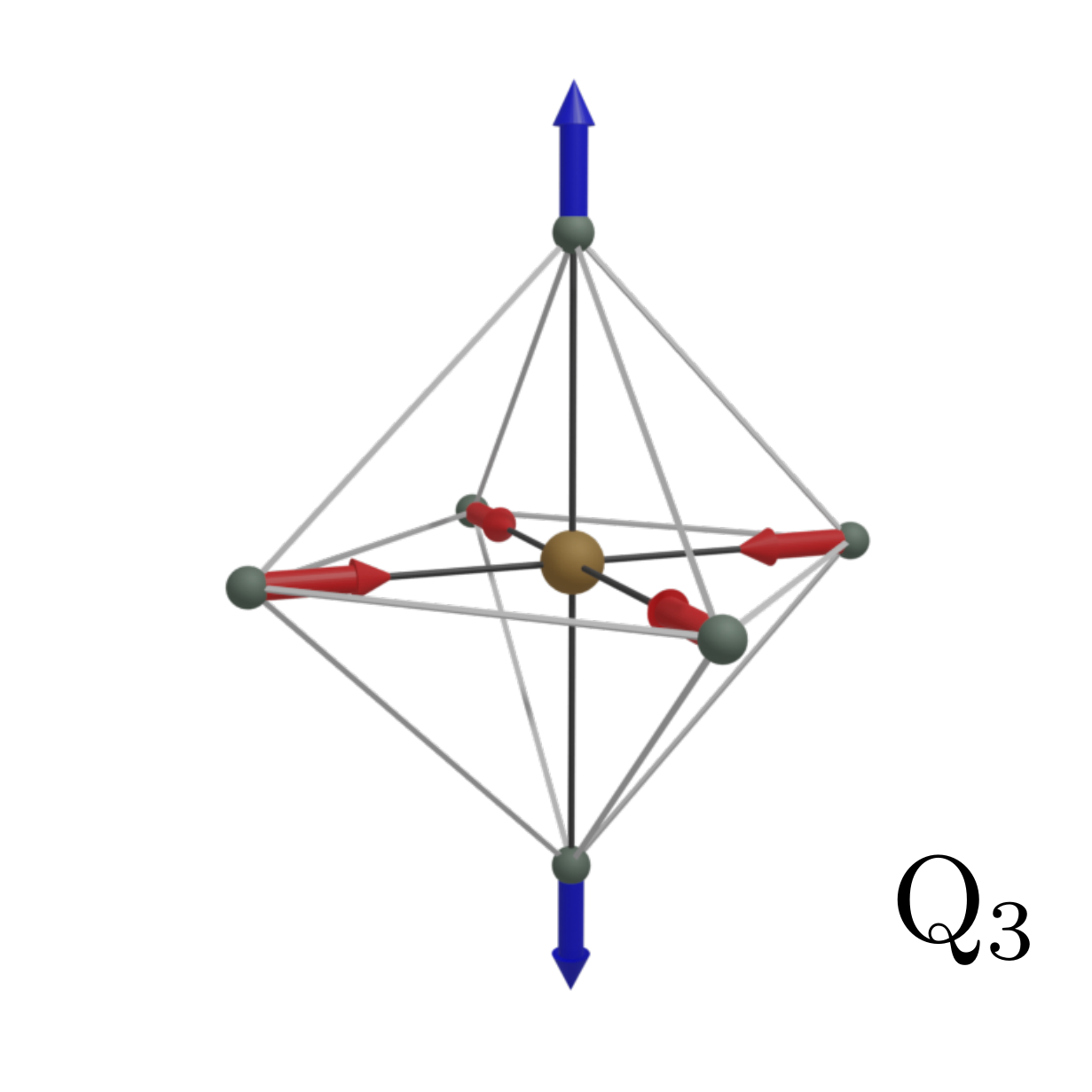}
    \end{subfigure}
    \hfill
    \begin{subfigure}{0.3\linewidth}
    \caption{}
    \centering
    \includegraphics[width=0.8\linewidth]{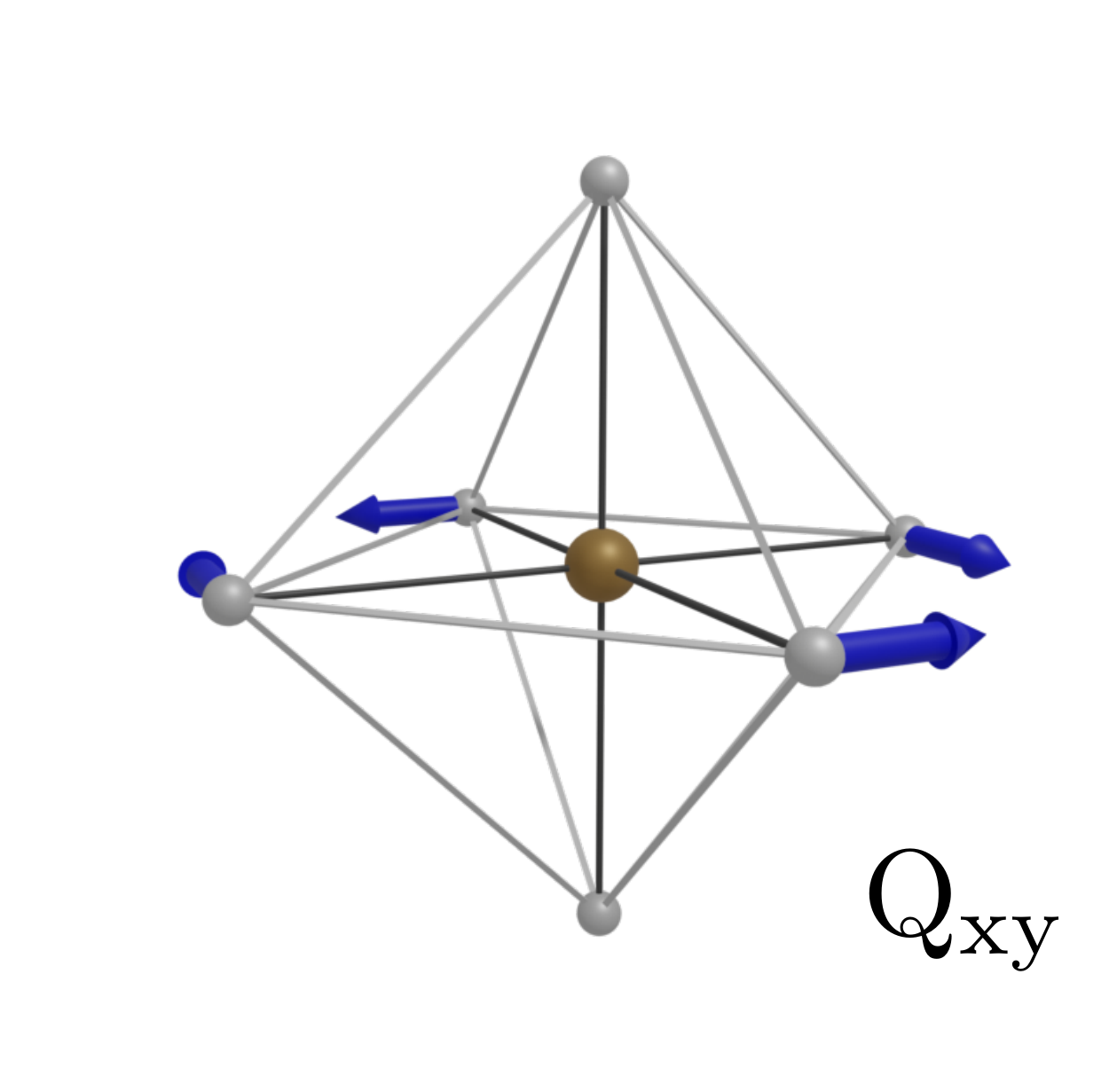}
    \end{subfigure}
    \caption{\textbf{Role of Jahn-Teller and SOC on  polaron stability}.  
    \textbf{(a)} Polaron energy $E_{pol}$ as a function of the SOC scaling factor $c_\lambda$ (For $c_\lambda=0$ SO if completely suppressed, whereas $c_\lambda$=1 refers to the full SO regime). $E_{pol}$ is progressively less negative for increasing SOC. 
    \textbf{(b)} JT tetragonal distortion amplitude $Q_3$ as a function of $c_\lambda$. The color scale and circle's size indicate the variation of $E_{pol}$ relative to the $c_\lambda=1$ case: SO rapidly quenches $Q_3$ and reduces the polaron stability. 
    \textbf{(c)} JT energy $E_{JT}$ at the polaron trapping site as a function of $c_\lambda$ for all three JT modes.  $Q_3$ is the only mode influenced by SOC.
    \textbf{(d-f)} Geometrical interpretation of the JT non-zero modes $Q_2$, $Q_3$ and $Q_{xy}$ respectively.}
    \label{fig:socjt}
\end{figure}

Although the $\mathrm{d^2}$ polaron possesses an intrinsic spin-orbital nature, SO coupling does not play in favour of polaron formation, as inferred from the progressive increase of the polaron energy $E_{pol}$ as a function of the SO strength 
shown in Fig.~\ref{fig:socjt}(a): the inclusion of SO destabilizes the polaron by about 80~meV. This behavior is linked to the effect of SO on the JT distortions recently elaborated by Streltsov and Khomskii, which suggests that for a $\mathrm{d^2}$ configuration, SO suppresses JT distortions~\cite{streltsov_interplay_2022, streltsov_jahn-teller_2020}. 
To shed light on this complex cross-coupling we have studied JT and polarons properties as a function of the effective SO coupling strength $\tilde{\lambda}=c_\lambda\lambda$ from $c_\lambda=0.1$ to  $c_\lambda=1$ (full SO). The resulting data are collected in Fig.~\ref{fig:socjt} and explained in the following.

Pristine $\mathrm{d^1}$ BNOO exhibits a cooperative JT ordering involving the $E_g$ modes $Q_2$ and $Q_3$~\cite{fiore_mosca_interplay_2021} as well as the trigonal $Q_{xy}$ mode (see Tab.~\ref{tabsm:qconfront}); these modes are graphically displayed in Fig.~\ref{fig:socjt}(d-f) and defined in Fig.~\ref{figsm:qs}. 
Upon charge trapping, the electrostatic potential of the OsO$_6$ octahedron increases due to the additional excess charge.
To counterbalance this energy cost, the oxygen cage expands according to the isotropic ($A_{1g}$) breathing-out mode $Q_1$. 
This expansion favours charge localization, and contributes $67\%$ of the total $E_{pol}$ (Fig.~\ref{figsm:polsta}), making $Q_1$ the major lattice contribution to polaron stability.
However, $Q_1$ is an isotropic deformation that does not break any local symmetry and therefore it is not related to the JT effect. Moreover, $Q_1$ is insensitive to SO and cannot play any role in the strong decrease of $E_{pol}$ with increasing $c_\lambda$ (see Fig.~\ref{figsm:activeqs}).
According to our data, only the tetragonal elongation along the $[001]$ axis $Q_3$ is strongly dependent on $c_\lambda$ (see Fig.~\ref{figsm:activeqs}).
In particular, Fig.~\ref{fig:socjt}(b) shows that increasing $c_\lambda$ yields a continuous decrease of $Q_3$, in agreement with the analysis of Ref.~\cite{streltsov_jahn-teller_2020}. 
Moreover, this SO-induced suppression of JT distortions is reflected on the JT energy (${E_{JT}}$), estimated from the potential energy
surface at different values of $c_\lambda$ and displayed in Fig.~\ref{fig:socjt}(c). 
Therefore, $Q_3$ appears to be the key JT mode explaining the coupling between SO and polaron stability, correlating the SO-induced decrease of $E_{pol}$ with the progressive quenching of $Q_3$ and associated reduction of ${E_{JT}}$.

This analysis provides a new conceptual framework to interpret the complex concerted interaction between JT, SO and polaron stability. Without SO the isotropic expansion $Q_1$ and the JT modes ($Q_2$, $Q_3$ and $Q_{xy}$) help polaron stabilization providing an energy gain $E_{JT}$ which depends on the distortion amplitude. SO dampens the JT distortion $Q_3$ leading to a reduction of $E_{JT}$ and consequentially a progressive increase of $E_{pol}$ (less stable polaron) with increasing SO coupling strength. This entangled \emph{spin-orbital Jahn-Teller bipolaron} 
develops in a relativistic background, is described by a spin-orbital $\mathrm{J_{eff}}={2}$ state and its stability is weakened by the SO-induced reduction of JT effects.

\begin{figure}[ht]
\centering
\includegraphics[width=0.95\linewidth]{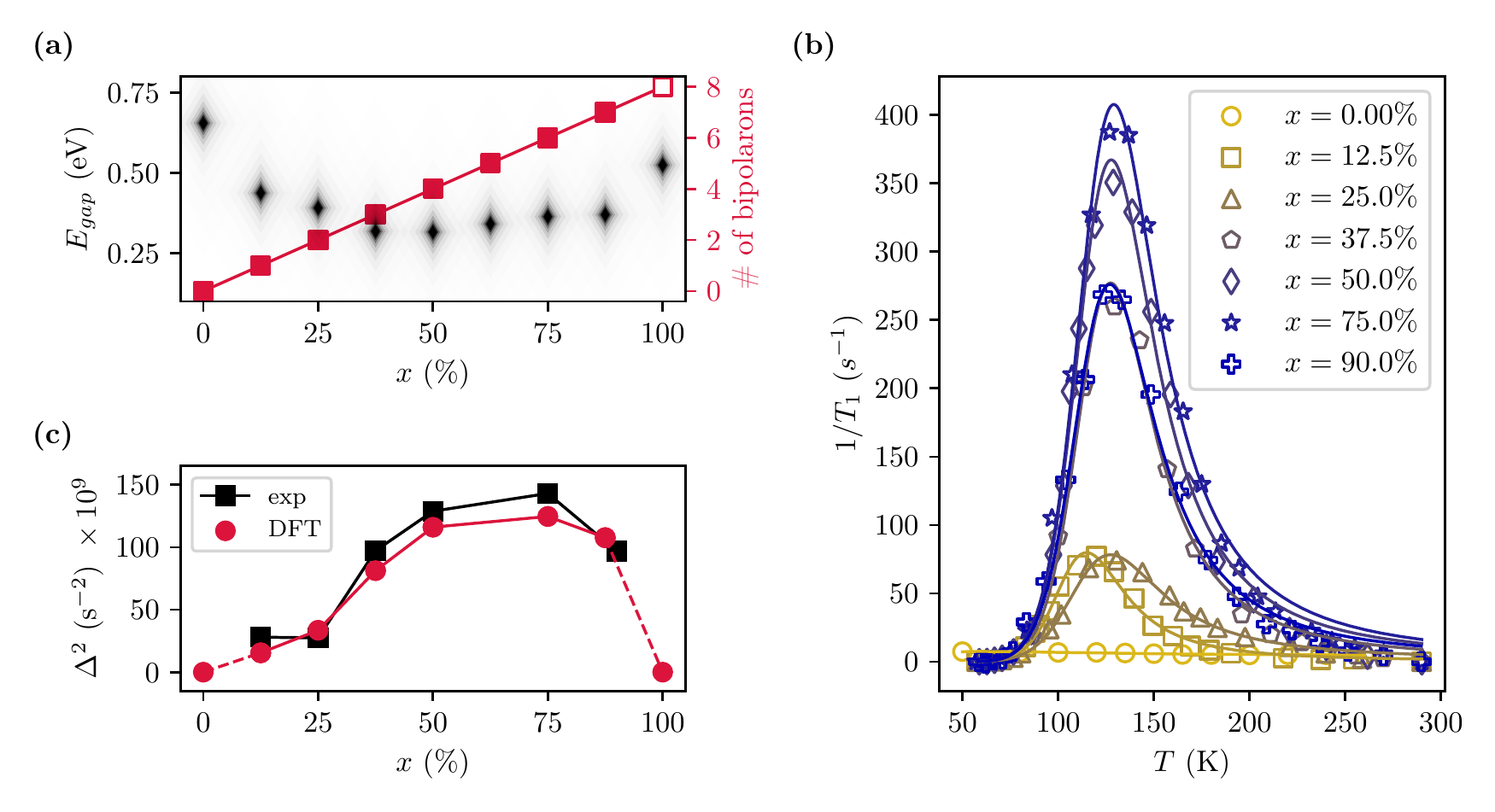}
\caption{\textbf{Polaron dynamics in $\mathbf{Ba_2Na_{1-x}Ca_xOsO6}$: DFT+NMR}.
\textbf{(a)} DFT energy gap (black diamonds) and number of bipolarons (red filled squares) as a function of doping. Chemically doped BNOO remains insulating for any doping concentration. The number of $\mathrm{d^2}$ bipolaron sites grows linearly with doping.
At full doping ($x=1$, corresponding to BCOO) all Os sites are  doubly occupied ($\mathrm{d^2}$) and polaron formation is completely quenched (empty square).
\textbf{(b)} Polaronic $1/T_1$ anomalous peak data and fitting curves calculated with the quadrupole relaxation model. For all investigated doping concentration doped BNOO exhibit a polaronic peak at approximately the same temperature $T$.
\textbf{(c)} Second moment $\Delta^2$ of the fluctuating field as extracted from the experimental data (black squares) compared with our predicted data obtained from the DFT polaron phonon field (red circles). The dashed lines connect to pristine BNOO ($x=0$) and BCOO ($x=1$) where the absence of polaron leads to $\Delta^2=0$. 
}
\label{fig:doping}
\end{figure}

Finally, we generalize our analysis to all doping concentrations disclosing the critical role of bipolarons in preserving the Mott state and elucidating the doping-induced modulation of the polaron phonon field as measured by NMR.
It is well established that a critical amount of carrier doping drives an insulator to metal transition (IMT)~\cite{RevModPhys.70.1039}. 
Formation of small polarons can delay the IMT, but at a critical polaron density, coalescence into a Fermi liquid prevails, leading to a metallic (or superconducting) phase~\cite{verdi2017,ciuchi,babio}. 
Notably, electron doped BNOO represents an exception: the insulating gap remains open at any concentration as shown in Fig.~\ref{fig:doping}(a) (the corresponding DOS are collected in Fig.~\ref{figsm:dosx}). 
This unique behaviour is explained by the absence of coherent hybridization between the bipolarons, facilitated by the large Os-Os distance of $\approx 5.9$~\AA\ in the double perovskite lattice. 
As illustrated in Fig.~\ref{fig:doping}(b), NMR shows the bipolaronic peak at $T_{P,1}\approx 130$~K at any Ca concentration with a virtually unchanged activation energy (see Tab.~\ref{smtab:fitresults}), confirming the DFT results which indicate a linear increase of number of $\mathrm{d^2}$ bipolarons with increasing doping (see Fig.~\ref{fig:doping}(a)).

The second moment of the fluctuating field $\Delta^2$ (see Eq.~(\ref{eq:bppmodel})) as a function of doping (Fig.~\ref{fig:doping}(c)) exhibits a dome shape, characterized by a progressive increase until $\approx 75\%$ Ca concentration, followed by a rapid decrease towards the full doping limit ($x=1$, BCOO) with all Os sites converted into a non-polaron $\mathrm{d^2}$ configuration in a undistorted and JT-quenched cubic lattice. 
To interpret these NMR measurements we have derived a model linking the polaron-induced spin-lattice relaxation rate $1/T_1$ 
with the oscillation of the polaron modes $\Delta Q_\xi  = \left< Q_\xi\right>_{d^2} - \left< Q_\xi\right>_{d^1}$ (with $\xi$ running over the dominant modes $Q_1$ and $Q_3$).
In particular, the average is taken over the distortions at the $\mathrm{d^2}$ and $\mathrm{d^1}$ sites obtained from DFT calculations.
The resulting compact formula reads (see the Methods section for a full derivation)
\begin{equation}
\Delta^2(x) = 
\frac{54(eQq_{ox})^2}
{5\hbar^2 R^8_{0}(x)} \left[ -\Delta Q_1^2(x) + 2 \Delta Q_3^2(x) \right] 
    \label{eq:modelmain}
\end{equation}
where $Q$ is the quadrupole moment of the $^{23}$Na nucleus, $q_{ox}$ is the charge of the oxygen ions (as obtained by DFT, 1.78$e$), 
and $R_0$ is the average Na-O bond length.
The obtained numerical data summarized
 in Fig.~\ref{fig:doping}(c) reproduce the experimental trend and indicate that, upon doping in $\mathrm{Ba_2Na_{1-x}Ca_xOsO_6}$, the main phonon contributions to polaron dynamics are encoded in the modulation of the breathing-out mode $Q_1$ and the tetragonal distortion $Q_3$ as a function of doping. 
In this regard, Fig.~\ref{fig:doping}(c) provides a transparent unprecedented microscopic interpretation of the polaron-driven spin-lattice relaxation rate $1/T_1$, here demonstrated for the quadrupolar polaron mechanism.

Summarizing, our study discloses a new type of polaron quasiparticle which is responsible for blocking the IMT even at ultrahigh doping and enables the coexistence of different spin-orbital $\mathrm{J_{eff}}$ states in the same compound.
This mixed-state can be interpreted as a precursor state towards the formation of the homogeneous $\mathrm{J_{eff}}=2$ state at full Na~$\to$~Ca substitution  (BCOO~\cite{khaliullin2021}), and the polaron is the main driving force of this transition.
In perspective, this work provides the conceptual means to explore polaron physics in quantum materials with strong spin–orbit coupling including topological~\cite{pesin_mott_2010}, Rashba~\cite{Manchon2015} and 2D materials~\cite{Ahn2020}, and pave the way for polaron spintronics~\cite{zutic_spintronics_2004}, polaron heavy-elements catalysis~\cite{D1TC02070F} and polaron multipolar magnetism~\cite{fiore_mosca_interplay_2021, PhysRevLett.127.237201, para2023, takayama21}.

\section*{Methods} 
\addcontentsline{toc}{section}{\protect\numberline{}Methods}
\label{sec:method}

\subsection*{Density Functional Theory}
\addcontentsline{toc}{subsection}{\protect\numberline{}Density Functional Theory}
\label{sec:methods_dft}

The electronic structure, structural deformations, and polaron hopping were studied using the fully relativistic version of \textsc{VASP}, employing the Perdew-Burke-Ernzerhof approximation for the exchange-correlation functional ~\cite{hobbs_fully_2000,liu_anisotropic_2015}.
All DFT calculations were performed with the magnetic moments' directions fixed to those of the low-temperature cAFM phase of BNOO~\cite{fiore_mosca_interplay_2021}. 
In addition, Dudarev's correction of DFT+U  was applied to account for strong electronic correlation effects, using a value of $U=3.4$~eV, which stabilizes the cAFM ordering in the pristine material. 
The computational unit cell is a $\sqrt{2}a\times \sqrt{2}a \times a$ supercell containing eight formula units, with $a = 8.27$~\AA\ referring to the lattice constant of the standard double perovskite unit cell, which contains four formula units. The sampling of the reciprocal space was done with a k-mesh of $4 \times 4 \times 6$, and an energy cutoff of 580~eV was selected for the plane wave expansion.
The SO contribution to the DFT energy functional could be manually controlled through a scaling parameter $c_\lambda$ using an in-house modified version of VASP.
The analysis of the JT effect was conducted using vibration modes defined by Bersuker~\cite{bersuker_jahn-teller_2006}, with some minor modifications, such as neglecting the rigid translation of the octahedra and including rigid rotations (see Sec.~\ref{secsm:jtstability}).
Electron doping was achieved by manually increasing the number of electrons in pristine BNOO. 
Charge neutrality is restored by adding a homogeneous background.
To extract polaronic energy levels and wavefunctions, a Wannier-like projection of Kohn-Sham wavefunctions was employed, using VASP non-collinear projected localized orbitals calculated on the polaronic Os site with the TRIQS's converter library~\cite{mosca_mott_2023} (see Sec.~\ref{secsm:dftlevels}).
DFT calculations for comparison with NMR data were performed on relaxed chemically doped $\sqrt{2}a\times \sqrt{2}a \times a$ supercells. 
The same number of k-points, energy cutoff and U were used as for the previous doping method. 

\subsection*{Dynamical Mean Field Theory}\addcontentsline{toc}{subsection}{\protect\numberline{}Dynamical Mean Field Theory}
\label{sec:methods_dmft}

For the analysis of the spin-orbital structure of the polaron levels we performed charge-self-consistent DFT+DMFT calculations within the Hubbard-I approximation \cite{hubbard_electron_1963, lichtenstein_ab_1998}, using \textsc{WIEN-2K}~\cite{blaha_wien2k_2020} and the TRIQS library \cite{parcollet_triqs_2015, aichhorn_triqsdfttools_2016}.
By using a $\sqrt{2}a\times \sqrt{2}a \times a$ supercell where one Na atom was substituted by one Ca ($12.5\%$ Ca concentration), we first calculated the polaronic ground state lattice structure, as explained in the previous \autoref{sec:methods_dft}, using VASP.
In DFT+HI calculations, the Wannier functions representing Os 5d states are constructed from the Kohn-Sham bands within the energy range $[-1,5]$~eV around the Kohn-Sham Fermi energy, which contains the Os $t_{2g}$ and most of the $e_{g}$ levels.
The fully-localized-limit double counting term on the polaronic Os site is set for the nominal $\mathrm{d^2}$ occupancy, as is appropriate for the quasi-atomic Hubbard-I approximation \cite{Pourovskii2007}, whereas for the rest of Os sites it is calculated for nominal $\mathrm{d^1}$.
The on-site interaction vertex for  the full $5d$ shell is specified by the parameters $U=3.5$~eV and $J_H=0.5$~eV, in agreement with the previous studies of pristine BNOO~\cite{fiore_mosca_interplay_2021}.

\subsection*{Nuclear Magnetic Resonance and muon Spin Rotation}
\addcontentsline{toc}{subsection}{\protect\numberline{}Nuclear Magnetic Resonance and muon Spin Rotation}
\label{sec:methods_exp}

We exploited the nuclear spin $I=3/2$ of $^{23}\mathrm{Na}$ nuclei in order to perform NMR spectroscopy on powder samples of $\mathrm{Ba_2Na_{1-x}Ca_{x}OsO_6}$, synthesised as described in Kesevan \emph{et al.}~\cite{kesavan_doping_2020}.
In particular, $^{23}\mathrm{Na}$ nuclei have a sizeable quadrupolar moment that allows to probe both magnetic and charge related dynamics.
We report spin-lattice ($1/T_1$) and spin-spin ($1/T_2$) relaxation rates as a function of temperature measured using an applied field of 7~T (details are reported in Sec.~\ref{secsm:dynamicalcons}).
We further analysed the anomalous peaks observed in these data by implanting a beam of polarised muons spin antiparallel to their momentum into the sample and applying a magnetic field of 10~mT and 100~mT parallel to the initial muon spin polarisation in order to measure the longitudinal muon relaxation rate $\lambda_\mu\equiv1/T_1^\mu$ (see Sec.~\ref{secsm:musr}).

\subsection*{Spin-lattice relaxation model}
\addcontentsline{toc}{subsection}{\protect\numberline{}Spin-lattice relaxation model}
\label{sec:methods_model}

The interaction of the nuclear quadrupole moment with an EFG $V_{\alpha\beta}$ can be written using spherical tensor operators $T_2^q$ as~\cite{mehring_principles_1983}
\begin{equation}
    H = \frac{eQ}{2I(2I-1)}\sum_{q=-2}^2 (-1)^q V_q T_2^{-q}
    \label{eq:quadH}
\end{equation}
where $Q$ is the quadrupole moment of the nucleus and $I=3/2$ is the nuclear spin.
To calculate the spherical component $V_q$ of the EFG we adopted a point-charge model of the NaO$_6$ octahedron.
The explicit expressions of the $V_q$'s are given in Sec.~\ref{secsm:model}.

The perturbation induced by fluctuations of the Na-O bonds resulting from polaron hopping is obtained from Eq.~(\ref{eq:quadH}) by expanding $V_q$ in terms of the bond variations $\Delta R_{i\alpha}(t)$
\begin{equation}
    H'(t) = \frac{eQ}{6} \sum_{i=1}^6 \sum_\alpha^{x,y,z} \sum_{qq'} \mathcal{D}^2_{q'q}(\mathcal{R}) T^{-q'}_2 w_{q}^{i\alpha} \Delta R_{i\alpha}(t)
    \label{eq:pert}
\end{equation}
where $w_{q}^{i\alpha}$ are the derivatives of the spherical components of the EFG with respect to the $\alpha$-th component of the $i$-th oxygen ion.
In Eq.~(\ref{eq:pert}) we have introduced the Wigner D-matrix $\mathcal{D}^2_{q'q}(\mathcal{R})$ to give account for the random orientation of the EFG reference frame with respect to the external magnetic field in powder samples.

The transition rate between two Zeeman levels $m$ and $m'$ averaged over all possible directions is given by
\begin{equation}
    W_{mm'} = \frac{(eQq_{ox})^2}{6\hbar^2} \sum_q \left| \bra{m} T_2^{-q} \ket{m'} \right|^2 \sum_{ij,\alpha\beta}  M^{(ij)}_{\alpha\beta} \int_{-\infty}^{\infty} \mathrm{d} t\ \overline{\Delta R_{i\alpha}(t) \Delta R_{j\beta}(0)}\ e^{-i\omega_{mm'}t}
    \label{eq:trans}
\end{equation}
where $q_{ox}$ is the oxygen ion charge in the point-charge model, $\omega_{mm'} = |\omega_m - \omega_{m'}|$ is the energy separation between the Zeeman levels, the matrices $M^{(ij)}_{\alpha\beta}$ are defined in Sec.~\ref{secsm:model} and $\overline{\Delta R_{i\alpha}(t) \Delta R_{j\beta}(0)}$ is the correlation function of the $\alpha$-th component of the $i$-th bond with the $\beta$-th of the $j$-th one.
To simplify the transition rate formula, some considerations on the crystal structure of $\mathrm{Ba_2Na_{1-x}Ca_xOsO_6}$ and small polaron dynamics are necessary. 

First, we notice that NaO$_6$ and OsO$_6$ octahedra are corner sharing in $\mathrm{Ba_2Na_{1-x}Ca_xOsO_6}$.
Therefore, the fluctuations of the $i$-th oxygen in the NaO$_6$ octahedron can be defined using the distortion modes $Q_\xi$ of the OsO$_6$ octahedron sharing the $i$-th oxygen with the NaO$_6$ one (see Fig.~\ref{figsm:qs}).
In this way the coordinate correlation functions can be expressed as correlation functions of the distortion modes $Q_{i,\xi}$ of the $i$-th OsO$_6$ octahedron.

Moving to polaron dynamics, if the $i$-th and the $j$-th octahedra are involved in an adiabatic hopping event within the time interval $ -\tau_c \lesssim t \lesssim \tau_c $ , in the LIS we have $Q_{i,\xi}(t) = Q_{j,\xi}(-t)$.
Thus, assuming that independent modes at the same site are uncorrelated, we can write all the correlation functions appearing in Eq.~(\ref{eq:trans}) as autocorrelation function of the independent distortion mode at each OsO$_6$ octahedron $\overline{Q_\xi(t) Q_\xi(0)}$.
To calculate these quantities, we recall that the MEHAM theory of adiabatic small polaron hopping relies on the classical treatment of phonon modes in the description of the site-jump process~\cite{holstein_studies_2000}.
In this limit, one can describe fluctuations using a phenomenological Langevin equation~\cite{feinberg_self-trapping_1986}.
Within this model, if the characteristic time of the fluctuations is much smaller than that of the hopping process ($\tau_c$), we speak of overdamped regime and the fluctuations' autocorrelations are given by
\begin{equation}
    \overline{Q_\xi(t) Q_\xi(0)} = \Delta Q^2_\xi\ e^{-|t|/\tau_c}
    \label{eq:corr}
\end{equation}
where $\Delta Q_\xi$ is the amplitude of the fluctuations of the $Q_\xi$ mode.
To validate this assumption, we evaluated the vibration frequencies $\omega_Q$ of the $Q_1$ and $Q_3$ oscillators from the potential energy curves obtained in the analysis of the Jahn-Teller modes (reported in Sec.~\ref{secsm:jtstability}).
We found them to be in the order of $\omega_Q \sim 10^{13}$~s$^{-1}$, as also reported in phonon spectra calculated by Voleti \emph{et al}.~\cite{para2023}. 
On the other hand, the correlation times extracted from NMR measurements are in the order of $\tau_c \sim 10^{-10}$~s, therefore $\omega_Q\tau_c \gg 1$, which corresponds to the overdamped regime~\cite{feinberg_self-trapping_1986}.
Moreover, we notice that the behaviour of the autocorrelations expressed in  Eq.~(\ref{eq:corr}) is commonly assumed in the description of spin-lattice relaxation processes~\cite{bloembergen_relaxation_1948, mehring_principles_1983, bonera_nuclear_1964}.

By neglecting correlation functions between opposite sites (Os-Os distance $\approx 8.29$~\AA), \emph{i.e.}~assuming only nearest-neighbour hopping (Os-Os distance $\approx 5.86$~\AA), the transition rate becomes
\begin{equation}
    W_{mm'} = \frac{3(eQq_{ox})^2}{2\hbar^2 R^8_0 } \sum_q \left| \bra{m} T^{q}_2 \ket{m'} \right|^2  \left( -\Delta Q_1^2 + 2 \Delta Q_3^2 \right) \frac{\tau_c}{1+(\omega_{mm'}\tau_c)^2}
    \label{eq:tr}
\end{equation}
where we have only considered fluctuations in the breathing-out mode $Q_1$ and the tetragonal mode $Q_3$ to be relevant, as deduced from the considerations expressed in the main text.

The spin-lattice relaxation time $T_1$ for a $I=3/2$ nucleus with quadrupolar interactions is given by~\cite{mehring_principles_1983}
\begin{equation}
    \frac{1}{T_1} = \frac{12}{5} ( W_1 + 4W_2 )
    \label{eq:1t1}
\end{equation}
where $W_1$ and $W_2$ are the transition rates for relaxations with selection rules $\Delta m=\pm 1$ and $\Delta m=\pm 2$ respectively.
By combining the latter Eq.~(\ref{eq:1t1}) with the transition rate formula in Eq.~(\ref{eq:tr}), we obtain the relation
\begin{equation}
    \frac{1}{T_1} = \Delta^2 \left[ \frac{\tau_c}{1+(\omega_0\tau_c)^2} + \frac{4\tau_c}{1+(2\omega_0\tau_c)^2} \right]
    \label{eq:fitmodel}
\end{equation}
which has been used to fit the NMR anomalous peak $1/T_1(T)$ with $\tau_c = \tau_0\exp(T_a/T)$, while $\Delta^2$ is given by
\begin{equation}
     \Delta^2 = \frac{54(eQq_{ox})^2}{5\hbar^2 R^8_0} \left( -\Delta Q_1^2 + 2 \Delta Q_3^2 \right) 
\end{equation}
which corresponds to Eq.~(\ref{eq:modelmain}) of the main text. Based on the above analysis and in analogy with the standard BPP model, $\Delta^2$ corresponds to the second moment of the fluctuating perturbation $H'(t)$ which is expressed in terms of the amplitude of the distortion modes.

\section*{Data availability}
The data that support the findings of this study are available from the corresponding author upon request.

\section*{Acknowledgements}
LC thank Michele Reticcioli and Luigi Ranalli for useful discussions. Support from the Austrian Science Fund (FWF) projects I4506 and J4698 is gratefully acknowledged. LC and DFM acknowledge the Vienna Doctoral School of Physics.
The computational results have been achieved using the Vienna Scientific Cluster (VSC). This work was supported in part by U.S. National Science Foundation (NSF) grant No. DMR-1905532 (V.F.M.), the NSF Graduate Research Fellowship under Grant No. 1644760 (E.G.), NSF Materials Research Science and Engineering Center (MRSEC) Grant No. DMR-2011876 (P.M.T. and P.M.W.). 
This work is based on experiments performed at the Swiss Muon Source SuS, Paul Scherrer Institute, Villigen, Switzerland.  

\subsection*{Author contributions}
C.F. conceived and supervised this project. L.C. executed all DFT calculations and analyzed the results, assisted by D. Fiore Mosca. L. V. Pourovskii conducted the DMFT calculations. 
S. Sanna and V. F. Mitrovi{\'c} have conceived and coordinated the experimental activity.
P. M. Tran and P. M. Woodward prepared the samples. G. Allodi, A. Tassetti, P. C. Forino, R. Cong and E. Garcia performed nuclear magnetic resonance measurements and analisys.
R. De Renzi, R. C., E. G. performed muon spin spectroscopy measurements and analisys. L.C. developed the spin-lattice relaxation model with inputs by C.F., G.A. and R.D.R.. C.F., L.C. and S.S. wrote the manuscript with input from all the authors.

\section*{Competing interests}
The authors declare no competing interests.

\printbibliography
\end{refsection}

\newpage
\setcounter{figure}{0}
\setcounter{equation}{0}
\renewcommand{\thefigure}{\arabic{figure}S}
\renewcommand{\thetable}{\arabic{table}S}
\renewcommand{\theequation}{\arabic{equation}S}
\renewcommand{\thesection}{\arabic{section}S}
\begin{refsection}

\begin{titlepage}
    \LARGE \textbf{Supplementary Informations}
\end{titlepage}

\newpage
\section{DFT polaronic levels}
\label{secsm:dftlevels}

The polaronic wavefunctions were calculated at the DFT level using the spinorial projected localised orbitals (PLOs) defined by Fiore Mosca~\emph{et al.}~\cite{mosca_mott_2023} and Sch\"uler~\emph{et al.}~\cite{schuler_charge_2018} within the projector augmented wave (PAW) scheme. 
Specifically, an effective Hamiltonian $H_{nn'}$ was obtained by projecting the Kohn-Sham energy levels $\epsilon_{m} (\mathbf{k})$ onto the subspace generated by the angular momentum $l=2$ wavefunctions centered at the polaronic site and radial part obtained from the PAW projectors, as implemented in the TRIQS dftTOOLs package~\cite{aichhorn_triqsdfttools_2016}.
The Hamiltonian was projected onto the local correlated space as obtained from the following equation, in which $H_{mm'}(\mathbf{k}) = \epsilon_{m} (\mathbf{k}) \delta_{mm'}$ 
\begin{equation}
 H^{\alpha \beta}_{nn'}= \sum_{\mathbf{k}}  \Tilde{P}^{\mathcal{C}\beta}_{nm'} (\mathbf{k})  H_{mm'} (\mathbf{k})  \big( \Tilde{P}^{\mathcal{C}\alpha}_{mn} (\mathbf{k}) \big)^* \ ,
\end{equation}
where
\begin{equation}
    P^{\alpha}_{\mathcal{C},L} (\mathbf{k}) = \sum_i \langle \chi_{L} | \phi_i \rangle \langle \Tilde{p}_i | \Tilde{\Psi}^{\alpha}_{ \mathbf{k}} \rangle \ ,
\end{equation}
are the spinorial PLOs written in the PAW formalism~\cite{mosca_mott_2023}. 

The projectors were calculated after convergence of the DFT self-consistent cycle within an energy window of $2.25\olddiv 9.75$~eV with respect to the Fermi level, in order to include all d levels of the polaronic Os site.
At this stage, the Hamiltonian was given in the $\ket{d_a, s_z}$ basis, where $a=\{xy, xz, yz, z^2, x^2 - y^2\}$ and $s_z = \pm 1/2$. 
In terms of $l^2$ and $l_z$ common eigenstates they are given by
\begin{align}
    &\ket{d_{xy}, s_z} = -\frac{i}{\sqrt{2}}\left( \ket{2, s_z} - \ket{-2, s_z} \right) \nonumber \\
    &\ket{d_{xz}, s_z} = -\frac{1}{\sqrt{2}}\left( \ket{1, s_z} - \ket{-1, s_z} \right) \nonumber \\
    &\ket{d_{yz}, s_z} = \frac{i}{\sqrt{2}}\left( \ket{1, s_z} + \ket{-1, s_z} \right) \\
    &\ket{d_{z^2}, s_z} = \ket{0, s_z}\nonumber \\
    &\ket{d_{x^2-y^2}, s_z} = \frac{1}{\sqrt{2}}\left( \ket{2, s_z} + \ket{-2, s_z} \right) \nonumber
\end{align}
The cubic harmonics $\ket{d_a}$ are defined with respect to the VASP internal reference frame, so that in our $\sqrt{2}a \times \sqrt{2}a \times a$ supercell the x, y and z axes correspond respectively to the $[110]$, $[1\Bar{1}0]$ and $[001]$ crystallographic axes. 
In order to have real harmonics defined in the reference frame of the $\mathrm{OsO_6}$ octahedron we rotate the orbital $l=2$ and spin $s=1/2$ subspaces using Wigner $D^j_{mm'}(\alpha, \beta, \gamma)$ matrices, defined as
\begin{equation}
    D^j_{mm'}(\alpha, \beta, \gamma) = \bra{j,m'} e^{-i\alpha J_z} e^{-i\beta J_y} e^{-i\gamma J_z}\ket{j,m}
\end{equation}
where $\alpha$, $\beta$ and $\gamma$ are Euler angles describing the reference frame rotation and $J_y$ and $J_z$ are cartesian coordinates of the angular momentum operator $\mathbf{J}$.
In our case the rotation is described by the angles $\alpha = 0$, $\beta = 0$, $\gamma = \pi/4$.

Since the crystal field interaction ($\sim 5$~eV) is much larger than SO interaction ( $\sim 0.3$~eV ) in BNOO, we can separate $t_{2g}$ and $e_g$ orbitals and consider only the former in our analysis.
In the octahedron reference frame they are defined as:
\begin{align}
    &\ket{t_{2g}^1}    = -\frac{1}{\sqrt{2}}( \ket{d_{xz}} + i\ket{d_{yz}} ) \nonumber \\
    &\ket{t_{2g}^0}    = \ket{d_{xy}} \\
    &\ket{t_{2g}^{-1}} = \frac{1}{\sqrt{2}}( \ket{d_{xz}} - i\ket{d_{yz}} ) \nonumber
\end{align}
One can easily show that the projection of the angular momentum operator $\mathbf{L}$ onto the $t_{2g}$ orbitals give
\begin{equation}
    P_{t_{2g}} \mathbf{L} P_{t_{2g}} = - \mathbf{l}
\end{equation}
where $\mathbf{l}$ is an effective angular momentum operator with $l=1$. 

\begin{figure}
    \centering
    \includegraphics[width=0.6\linewidth]{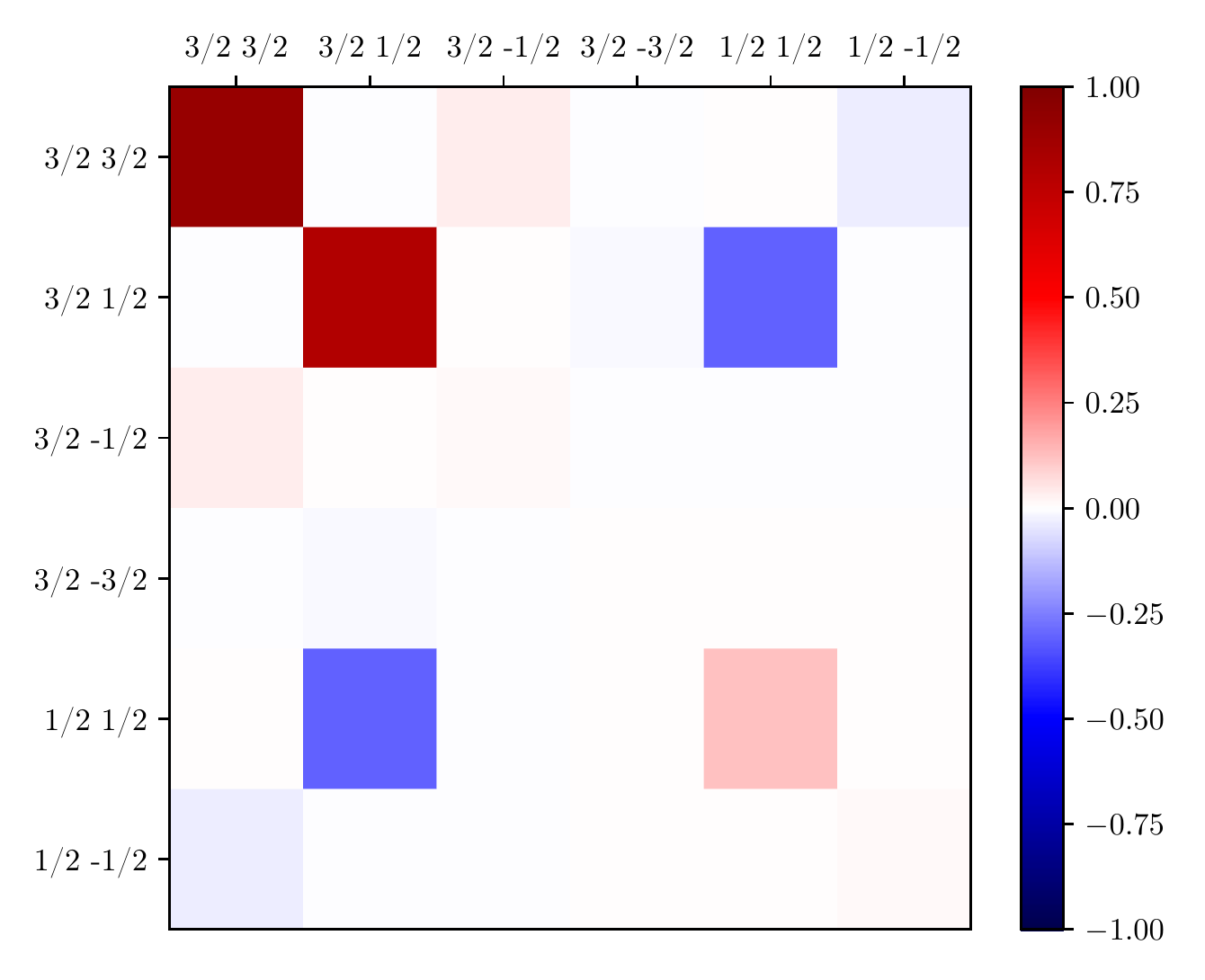}
    \caption{Diagram of the polaronic occupation matrix as calculated from DFT non-collinear projectors. Axes are labelled according to $(j, j_z)$ values. Colors represent the value of the occupation matrix reported in Tab.~\ref{tabsm:occmat}. Dark red on the diagonal corresponds to level occupation equal to one. The diagram shows that the DFT polaron occupies $J_{\text{eff}}=3/2$ states $\ket{3/2, 3/2}$ and $\ket{3/2,1/2}$ with the latter state slightly mixed with the $J_{\text{eff}}=1/2$ $\ket{1/2,1/2}$.}
    \label{figsm:colorocc}
\end{figure}

At this point, we employ Clebsch-Gordan coefficients to construct $\mathrm{J_{eff}}=3/2$ and $\mathrm{J_{eff}}=1/2$ states out of the $t_{2g}$ ones.
To conclude, we rotated again our basis with Wigner $D$-matrices to bring the angular momentum quantization axis along the direction of magnetisation, which lies in the $(110)$ plane.
The occupation matrix thus obtained is pictorially reproduced in Fig.~\ref{figsm:colorocc} and numerically in Tab.~\ref{tabsm:occmat}.

\begin{table}
\centering
\begin{tabular}{c|cccccc}
          & 3/2, 3/2 & 3/2, 1/2 & 3/2, -1/2 & 3/2, -3/2 & 1/2, 1/2 & 1/2, -1/2 \\ \hline
3/2, 3/2  & 0.90     & 0.00     & 0.04      & 0.00      & 0.00     & -0.04     \\
3/2, 1/2  & 0.00     & 0.80     & 0.00      & -0.01     & -0.31    & 0.00      \\
3/2, -1/2 & 0.04     & 0.00     & 0.01      & 0.00      & 0.00     & 0.00      \\
3/2, -3/2 & 0.00     & -0.01    & 0.00      & 0.00      & 0.00     & 0.00      \\
1/2, 1/2  & 0.00     & -0.31    & 0.00      & 0.00      & 0.12     & 0.00       \\
1/2, -1/2 & -0.04    & 0.00     & 0.00      & 0.00      & 0.00     & 0.01     \\ \hline
\end{tabular}
\caption{Polaronic occupation matrix calculated from DFT non-collinear projectors. Column and row labels indicate the matrix components on the $J_{\text{eff}}=3/2$ and $J_{\text{eff}}=1/2$ levels.}
\label{tabsm:occmat}
\end{table}

\clearpage
\section{HI polaronic levels}
\label{secsm:hi}

\begin{figure}[ht]
\begin{minipage}{0.25\linewidth}
    \centering
    \includegraphics[width=\linewidth]{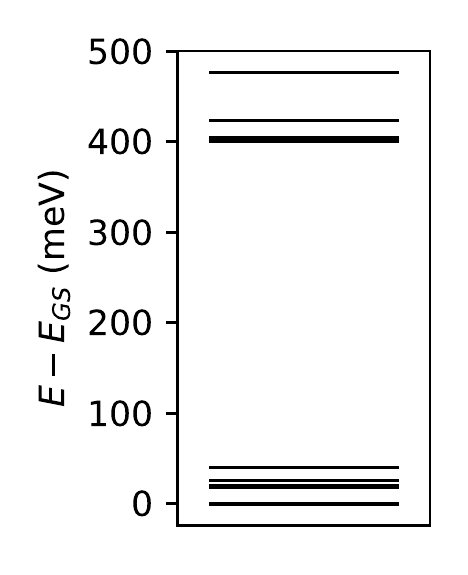}
\end{minipage}
\hfill
\begin{minipage}{0.7\linewidth}
    \centering
    \begin{tabular}{ll}
    \toprule
    energy (meV) & state \\
    \midrule
       0.00 & $+0.999|2;0\rangle-( 0.031-0.003 i)|2;2\rangle-( 0.031+0.003 i)|2;-2\rangle$  \\ 
       18.6 & $+0.707|2;-1\rangle-( 0.701-0.094 i)|2;1\rangle$  \\ 
       20.3 & $+0.707|2;1\rangle+( 0.701+0.094 i)|2;-1\rangle$  \\
       25.6 & $+0.706|2;2\rangle + 0.706|2;-2\rangle + 0.044|3/2;1/2\rangle$  \\
       40.2 & $+0.707|2;-2\rangle - 0.707|2;2\rangle+( 0.000 -0.004 i)|2;0\rangle$  \\
       \bottomrule
    \end{tabular}
\end{minipage}
\caption{Lowest two-electron energy levels (left) and wavefunctions of the first five levels. The first row of the table indicates that the polaronic ground state has a spin-orbital character with $J_{\text{eff}}=2$.}
\label{figsm:2ellev}
\end{figure}

Charge self-consistent DFT+DMFT calculations (see \autoref{sec:methods_dmft} of the main text) predicted a multiplet structure of the two-electron levels at the polaronic site, as depicted in the left panel of Fig.~\ref{figsm:2ellev}.
In particular, we can distinguish a low lying quintuplet, a triplet about 0.4~eV above the ground state multiplet and a singlet.
The next excited state lies about 1~eV above those represented in Fig.~\ref{figsm:2ellev}.
By looking at the wavefunctions of the ground state quintuplet, we can see that they are well represented by $\mathrm{J_{eff}}=2$ spin-orbital states (right panel of Fig.~\ref{figsm:2ellev}.

\begin{figure}[ht]
    \centering
    \includegraphics[width=0.7\linewidth]{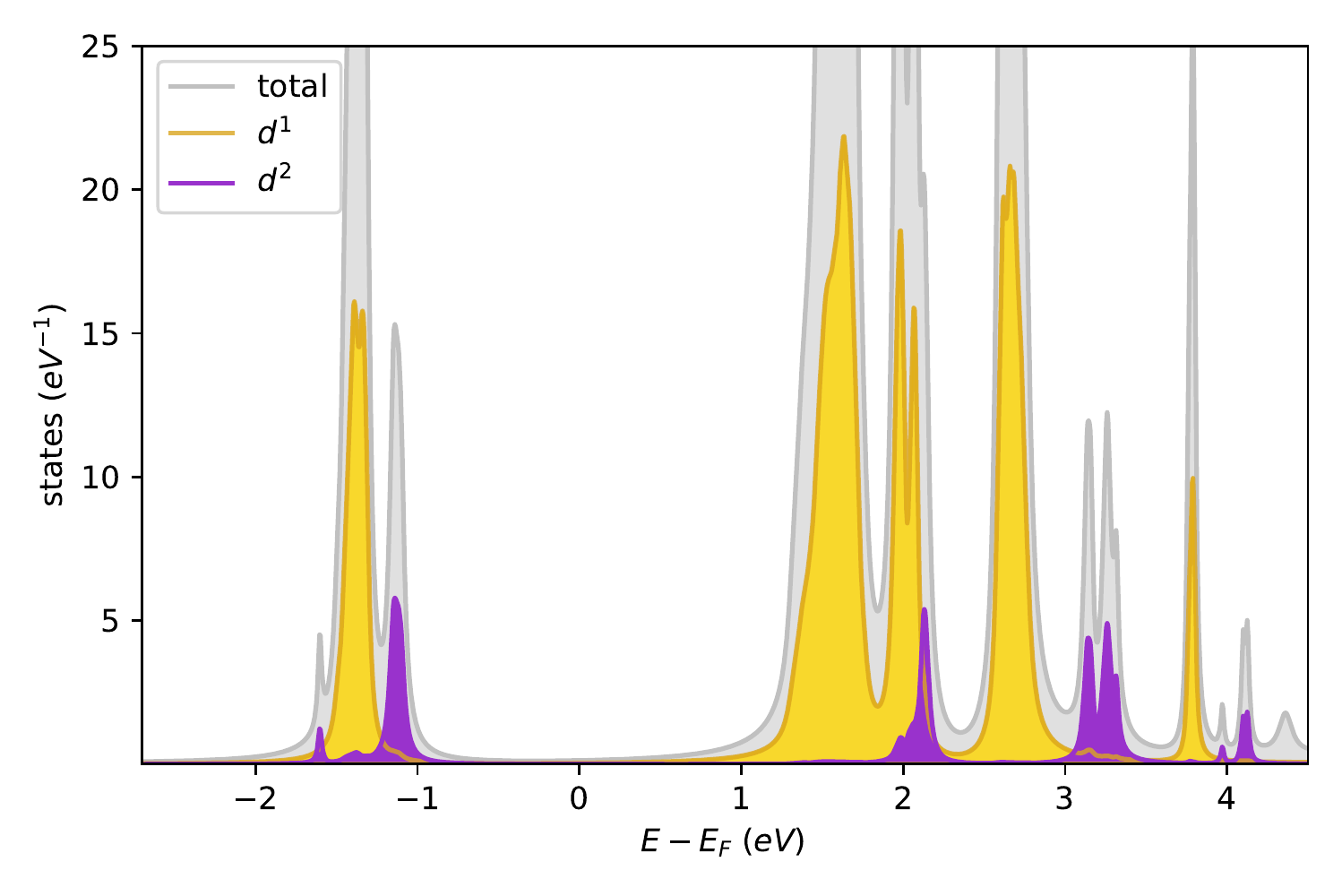}
    \caption{Single particle spectral density. Contributions of the $\mathrm{d^1}$ and $\mathrm{d^2}$ sites are highlighted respectively in yellow and violet. The violet polaronic peak lies just above the conduction band, around $-1$~eV.}
    \label{figsm:hidos}
\end{figure}

The one-electron spectral density was also calculated for the $\mathrm{d^1}$ and $\mathrm{d^2}$ states and is reported in Fig.~\ref{figsm:hidos}.
The DFT+HI polaronic states appear in the band gap just above the valence $\mathrm{d^1}$ states.

\clearpage
\section{Polaron formation energy vs JT modes}
\label{secsm:jtstability}

\begin{figure}[ht]
\begin{minipage}{0.49\linewidth}
    \centering
    \includegraphics[width=\linewidth]{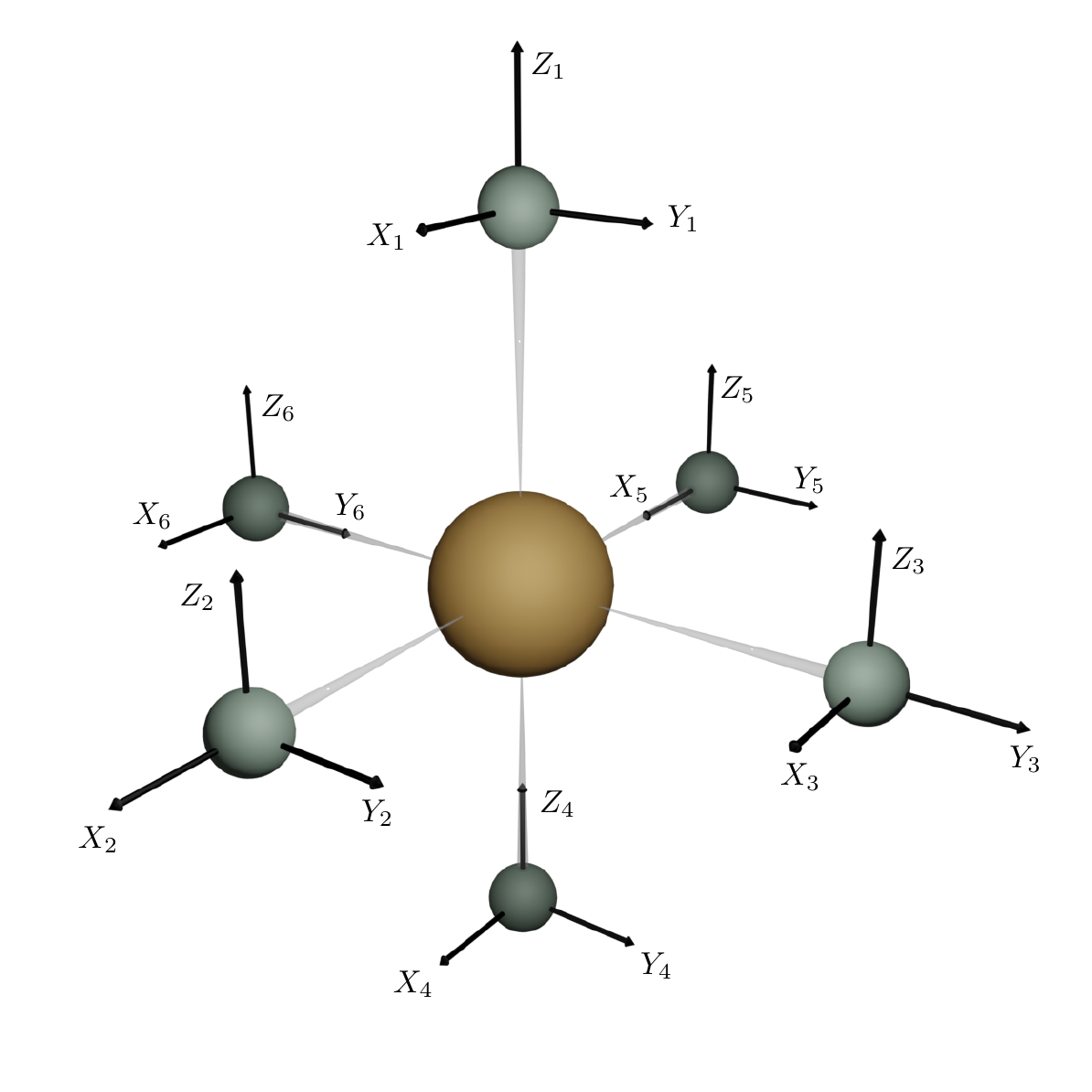}
\end{minipage}
\hfill
\begin{minipage}{0.49\linewidth}
\centering
\begin{tabular}{lll}
\toprule
$A_{1g}$                  & $Q_1$      & $(X_2-X_5+Y_3-Y_6+Z_1-Z_4)/\sqrt{6}$    \\\\
\multirow{2}*{$E_g$}      & $Q_2$      & $(X_2-X_5-Y_3+Y_6)/2$                   \\
                          & $Q_3$      & $(2Z_1-2Z_4-X_2+X_5-Y_3+Y_6)/2\sqrt{3}$ \\\\
\multirow{3}*{$T_{2g}$}   & $Q_{yz}$   & $(Z_3-Z_6+Y_1-Y_4)/2$                   \\
                          & $Q_{xz}$   & $(X_1-X_4+Z_2-Z_5)/2$                   \\
                          & $Q_{xy}$   & $(Y_2-Y_5+X_3-X_6)/2$                   \\\\
\multirow{3}*{$T'_{1u}$}  & $Q_{x}'$   & $(X_1+X_3+X_4+X_6)/2$                   \\
                          & $Q_{y}'$   & $(Y_1+Y_2+Y_4+Y_5)/2$                   \\
                          & $Q_{z}'$   & $(Z_1+Z_3+Z_4+Z_6)/2$                   \\\\
\multirow{3}*{$T''_{1u}$} & $Q_{x}''$  & $(X_2+X_5)/\sqrt{2}$                    \\
                          & $Q_{y}''$  & $(Y_3+Y_6)/\sqrt{2}$                    \\
                          & $Q_{z}''$  & $(Z_1+Z_4)/\sqrt{2}$                    \\\\
\multirow{3}*{$T_{1u}$}   & $Q_{xy}'$  & $(X_3+X_6-X_1-X_4)/2$                   \\
                          & $Q_{yz}'$  & $(Y_1+Y_4-Y_2-Y_5)/2$                   \\
                          & $Q_{xz}'$  & $(Z_2+Z_5-Z_3-Z_6)/2$                   \\\\
\multirow{3}*{rot.}       & $Q_{r,yz}$ & $(Z_3-Z_6+Y_4-Y_1)/2$                   \\
                          & $Q_{r,xz}$ & $(X_1-X_4-Z_2+Z_5)/2$                   \\ 
                          & $Q_{r,xy}$ & $(Y_2-Y_5-X_3+X_6)/2$                   \\
\bottomrule
\end{tabular}%
\end{minipage}
\caption{Definition of the octahedron vibration coordinates $Q_i$ as a function of the cartesian distortions $X_i$, $Y_i$ and $Z_i$.}
\label{figsm:qs}
\end{figure}

The JT normal coordinates used in our analysis are defined as linear combinations of the Cartesian displacements $X_i$, $Y_i$, and $Z_i$ of the oxygen atoms located at the corners of the distorted octahedron structure (see Fig.~\ref{figsm:qs}). 
These displacements are defined with respect to their corresponding positions in a reference undistorted structure, denoted by $x_i^{(0)}, y_i^{(0)}, z_i^{(0)}$. 
Starting from $X_i$, $Y_i$, and $Z_i$, JT normal coordinates $Q_i$ can be defined that transform according to the irreducible representations of the octahedral group $O_h$~\cite{bersuker_jahn-teller_2006}, as represented in Fig.~\ref{figsm:qs}.

We used the BNOO cubic structure with full $Fm\Bar{3}m$ symmetry~\cite{stitzer_crystal_2002} as a reference.
Cartesian directions $X_i$, $Y_i$ and $Z_i$ correspond to crystallographic axes $[100]$, $[010]$ and $[001]$ respectively.
In all our calculations, we find only four modes to be different from zero: $Q_1$, $Q_2$, $Q_3$, and $Q_{xy}$.
Their values for both the pristine BNOO with cAFM ordering and the polaronic site are reported in Tab.~\ref{tabsm:qconfront}.
The $T_{2g}$ mode $Q_{xy}$ is not mentioned in previous studies on pristine BNOO~\cite{fiore_mosca_interplay_2021}.
\begin{table}[ht]
\centering
\begin{tabular}{ccccc}
         & $Q_1$ (\AA) & $Q_2$ (\AA) & $Q_3$ (\AA) & $Q_{xy}$ (\AA) \\ \hline
pristine & 0.057                      & 0.017                      & -0.005                     & 0.019                         \\
polaron  & 0.138                      & -0.011                     & 0.014                      & -0.010                        \\ \hline
\end{tabular}
\caption{Non-zero deformation modes for an $\mathrm{OsO_6}$ octahedron in the pristine material (first row) and for a polaronic one (second row).}
\label{tabsm:qconfront}
\end{table}

\begin{figure}[ht]
    \centering
    \includegraphics[width=\linewidth]{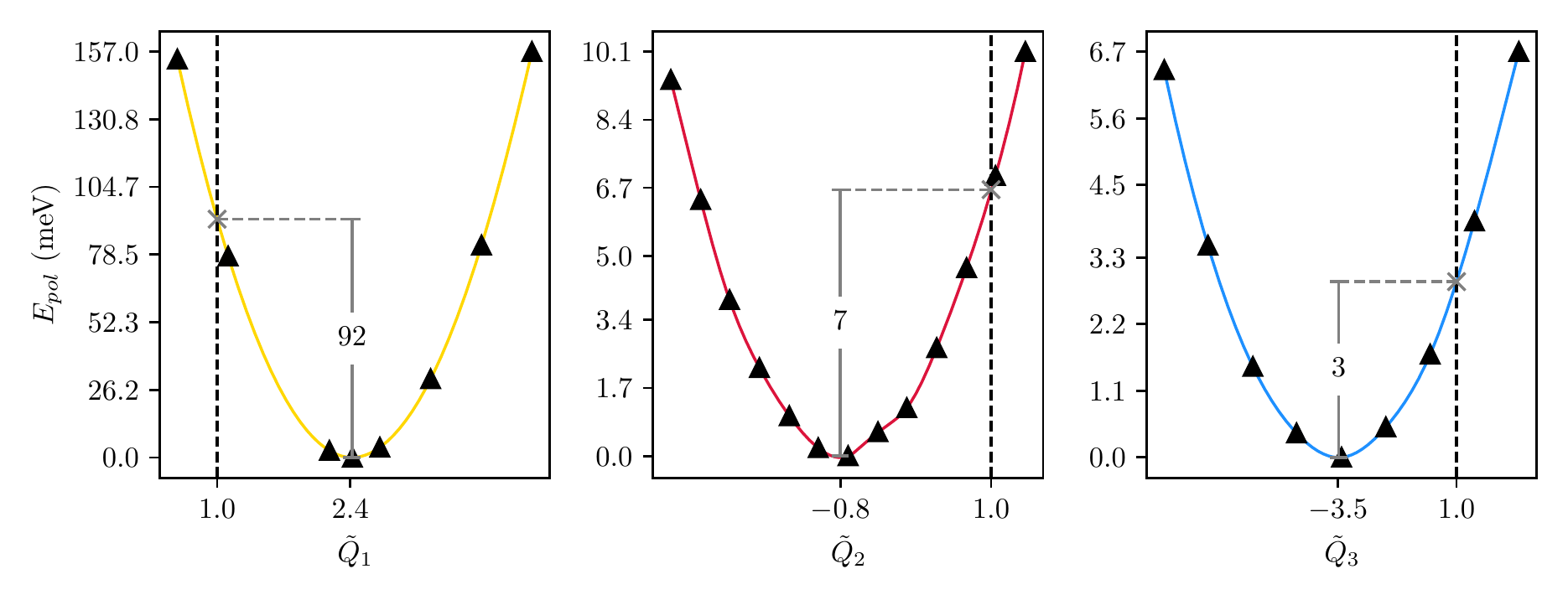}
    \caption{Polaron energy $E_{pol}$ as function of the normalised $\Tilde{Q}_i$ modes for $c_{\lambda}=1$. Dashed lines correspond to the pristine values of $\Tilde{Q}_i$. 
    Solid grey lines indicate the lattice energy gain given by a polaronic distortion with respect to the pristine one. The breathing-out mode $\Tilde{Q}_1$ presents the major contribution amounting to $\approx 67\%$ of the polaron energy.}
    \label{figsm:polsta}
\end{figure}

We provide in the following the analysis of the effect of non-zero modes on the polaronic formation energy and their relation to SO effects. 
First, we observe how polaron energy $E_{pol}$ changes as a function of the isotropic mode $Q_1$ and the $E_g$ modes $Q_2$ and $Q_3$. 
In particular, starting from the distortions obtained for the polaronic ground state, we calculate all the $Q_i$ of Fig.~\ref{figsm:qs}. 
Then, we fix all $Q_i$ but the one we want to study and invert the transformation between $Q_i$ and Cartesian displacements to generate the new structure, where the polaron has one mode changed and all others are left unchanged. 
By fixing the position of the polaronic $\mathrm{OsO}_6$ octahedron and letting all other ions to relax, we calculate the polaron energy always with respect to the same delocalised configuration. 
In this way, we obtain the parabolas reported in Fig.~\ref{figsm:polsta}. 
Here, the normalised distortions $\Tilde{Q}_i$ are defined as the ratio of the polaronic and pristine distortion modes.
In this way we can easily estimate the energy gained by the polaron by change one mode $Q_i$ with respect to its value in the pristine case $\Tilde{Q}_i = 1$.
The same calculations for $Q_{xy}$ show an energy gain of the same order of magnitude of those obtained for the $E_{g}$  modes.

\begin{figure}[ht]
    \centering
    \includegraphics[width=\linewidth]{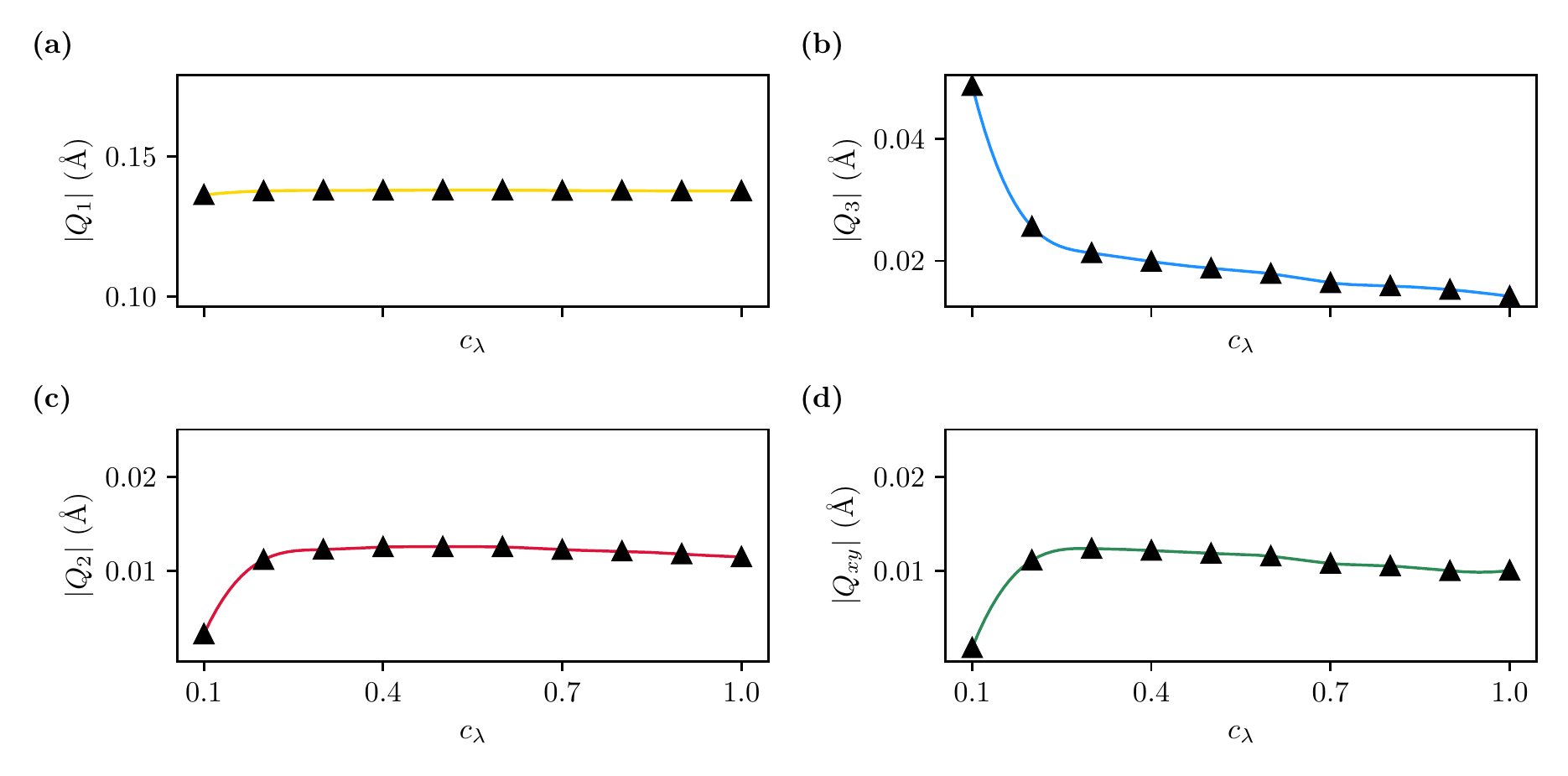}
    \caption{Non-zero deformation modes behaviour as a function of SO coupling strength $c_\lambda$. \textbf{(a)} the breathing-out mode $Q_1$ shows no SO dependence. \textbf{(b)} the tetragonal mode $Q_3$ is quenched by SOC. \textbf{(c-d)} the orthorhombic mode $Q_2$ and the trigonal one $Q_{xy}$ are completely suppressed at very low SO and show a plateau for $c_\lambda \gtrsim 0.2$.}
    \label{figsm:activeqs}
\end{figure}

As a second step, we explore the relation between the JT modes and SO intensity $c_\lambda$.
The $A_{1g}$ mode $Q_1$, corresponding to an isotropic expansion of the octahedron, stays constant throughout the range of SO intensity (see Fig.~\ref{figsm:activeqs}(a)).
For the $E_{g}$ and $T_{2g}$ modes, two different behaviours could be distinguished.
The tetragonal elongation in the $[001]$ direction $Q_3$ decreases with increasing $c_\lambda$, as already mentioned in the main text and also found by Streltsov and Khomskii~\cite{streltsov_jahn-teller_2020}.
On the other hand, $Q_2$ and $Q_{xy}$ remain constant above $c_\lambda \simeq 0.2$ and rapidly go to zero for $c_\lambda < 0.2$, as represented in Fig.~\ref{figsm:activeqs}(c-d). 
However, previous studies predicted exactly the opposite for a $d^2$ JT impurity: all $E_g$ and $T_{2g}$ distortions should decrease in amplitude for increasing $c_\lambda$~\cite{streltsov_jahn-teller_2020, streltsov_interplay_2022}.
Moreover, the different behaviours of $Q_2$ and $Q_3$ suggest
Therefore, we supposed that the $Q_2$ and $Q_{xy}$ distortions have a different origin than the $Q_3$ one.

To prove this assumption, we estimate the JT energy $E_{JT}$ of the polaronic octahedron as a function of $c_\lambda$ in a quasimolecular approximation~\cite{bersuker_jahn-teller_2006,iwahara_spin-orbital-lattice_2018}.
In particular, we calculate all the $Q_i$ from the polaronic ground state structure, fix all of them but a chosen one, and invert the transformation of Fig.~\ref{figsm:qs} between Cartesian displacements and generalised modes $Q_i$ to generate the distorted structures needed to construct the ionic potential energy surface. 
For each distorted structure, total energies are calculated with the occupation matrix constrained to that of the polaronic ground state~\cite{allen_occupation_2014}. 
The parabolas $E(Q_i)$ obtained in this way have been fitted to estimate the JT energy gain $E_{\mathrm{JT}}$ as the difference between the energy at the minimum and that corresponding to the structure with the varying $Q_i$ equal to zero:
\begin{equation}
    E_{\mathrm{JT}} = E( Q_i^{pol} ) - E(Q_i = 0)
\end{equation}
where $Q_i^{pol}$ is the value of the distortion for the polaronic ground state.
The results are shown in Fig.~\ref{fig:socjt} of the main text. 
While $E_{\mathrm{JT}}$ was one order of magnitude more negative for the tetragonal distortion $Q_3$ at weak SO coupling, for $Q_2$ and $Q_{xy}$ the JT energy gain remains smaller than $1$~meV throughout the explored $c_\lambda$ range. 
We therefore assume that their contribution to the polaron stability is negligible compared to the $Q_3$ one.

\clearpage
\section{DOS for different Ca concentrations}
\label{secsm:dosx}

\begin{figure}[ht]
\centering
\begin{subfigure}{0.32\linewidth}
    \includegraphics[width=\linewidth]{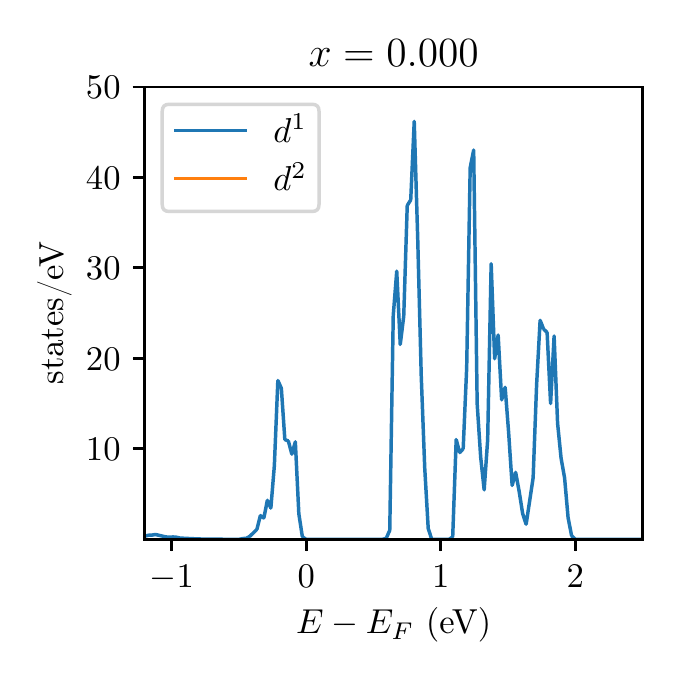}
\end{subfigure}
\hfill
\begin{subfigure}{0.32\linewidth}
    \includegraphics[width=\linewidth]{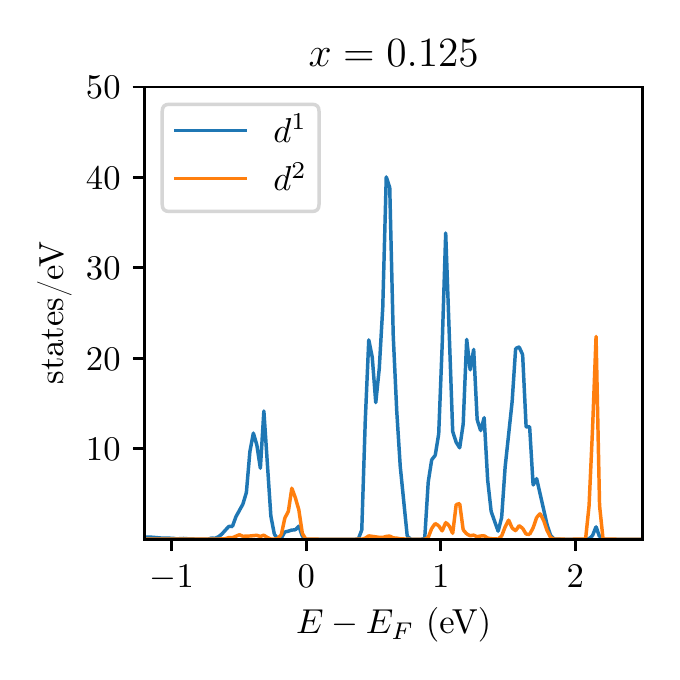}
\end{subfigure}
\hfill
\begin{subfigure}{0.32\linewidth}
    \includegraphics[width=\linewidth]{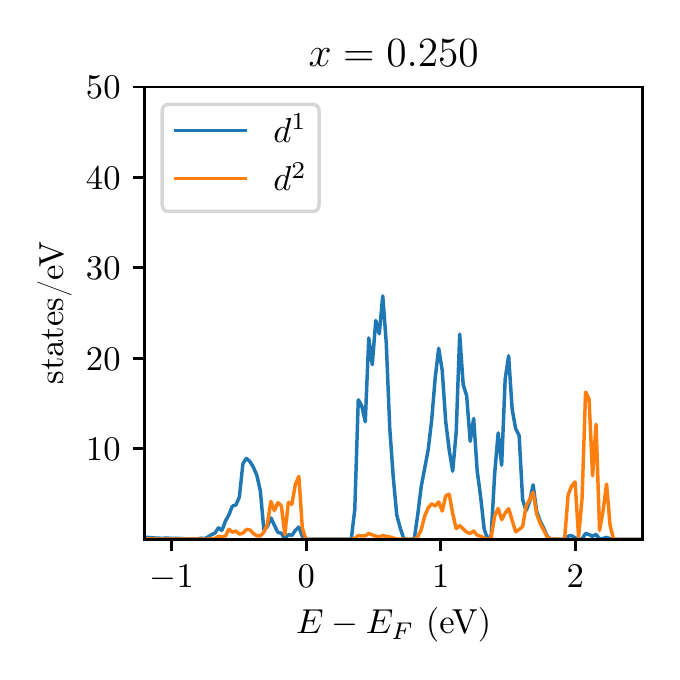}
\end{subfigure}
\vfill
\begin{subfigure}{0.32\linewidth}
    \includegraphics[width=\linewidth]{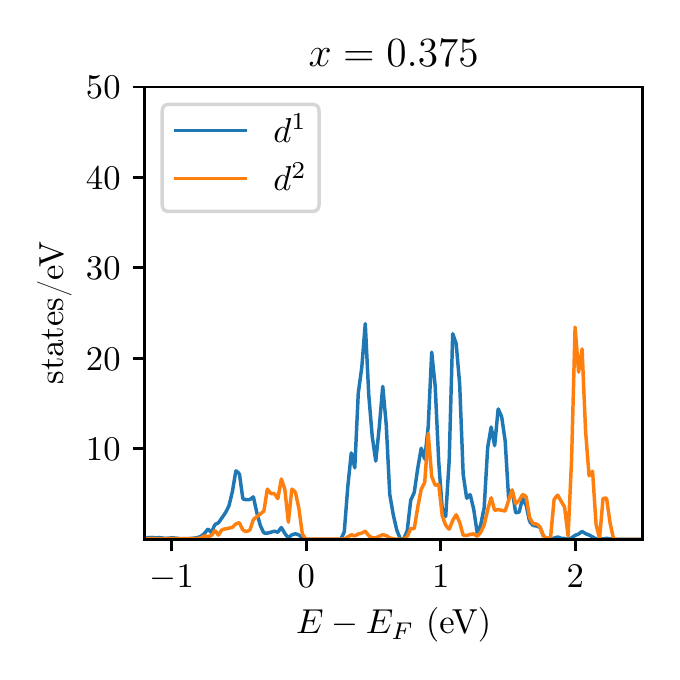}
\end{subfigure}
\hfill
\begin{subfigure}{0.32\linewidth}
    \includegraphics[width=\linewidth]{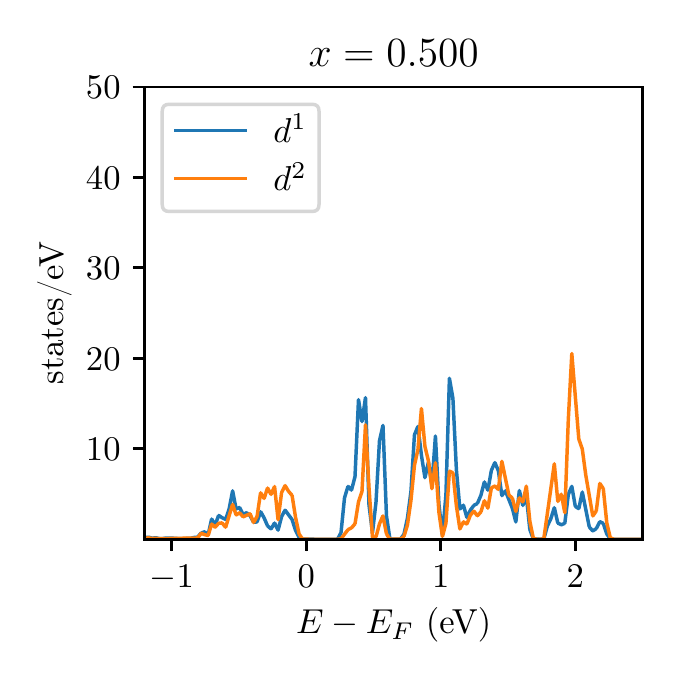}
\end{subfigure}
\hfill
\begin{subfigure}{0.32\linewidth}
    \includegraphics[width=\linewidth]{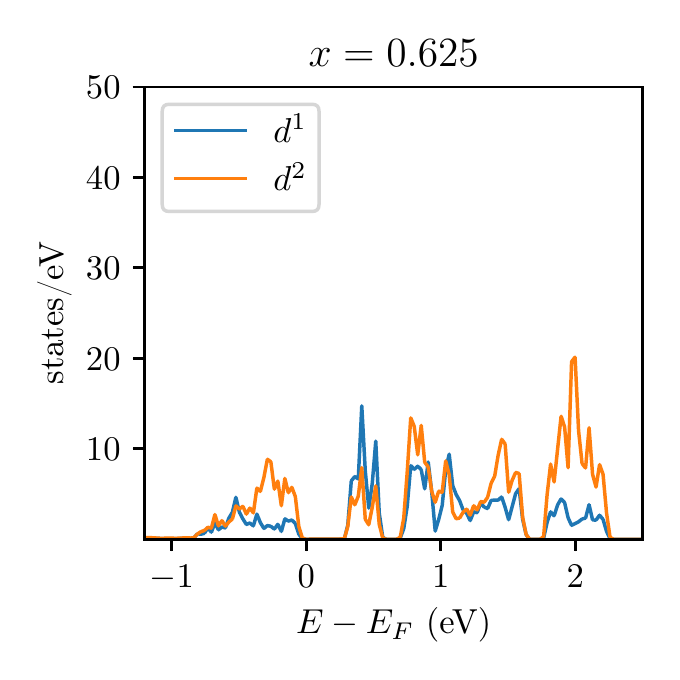}
\end{subfigure}
\vfill
\begin{subfigure}{0.32\linewidth}
    \includegraphics[width=\linewidth]{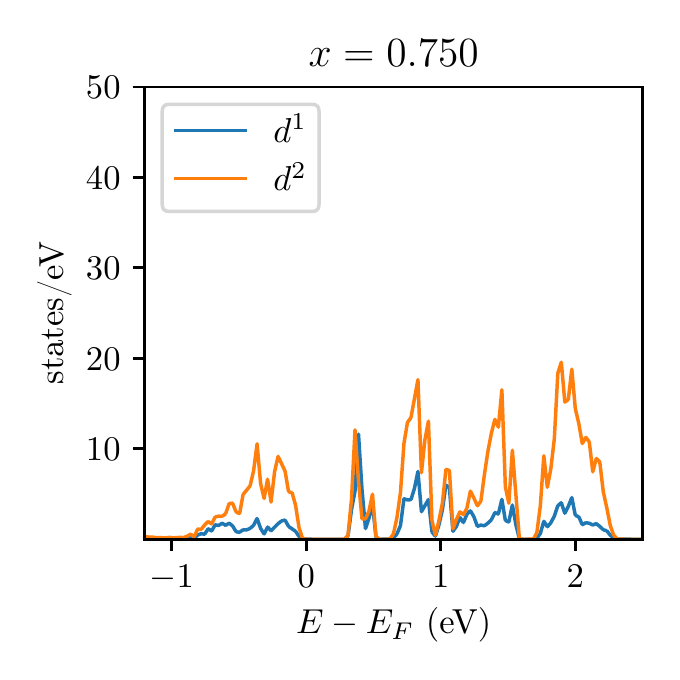}
\end{subfigure}
\hfill
\begin{subfigure}{0.32\linewidth}
    \includegraphics[width=\linewidth]{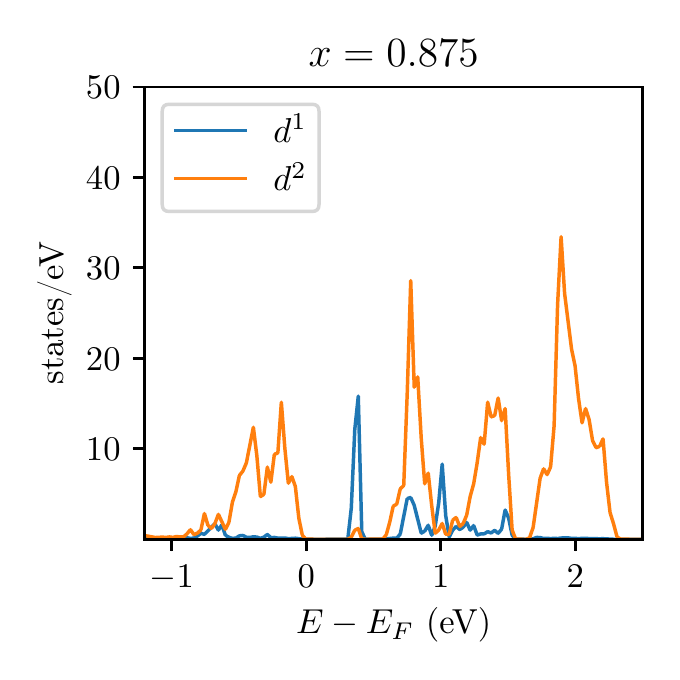}
\end{subfigure}
\hfill
\begin{subfigure}{0.32\linewidth}
    \includegraphics[width=\linewidth]{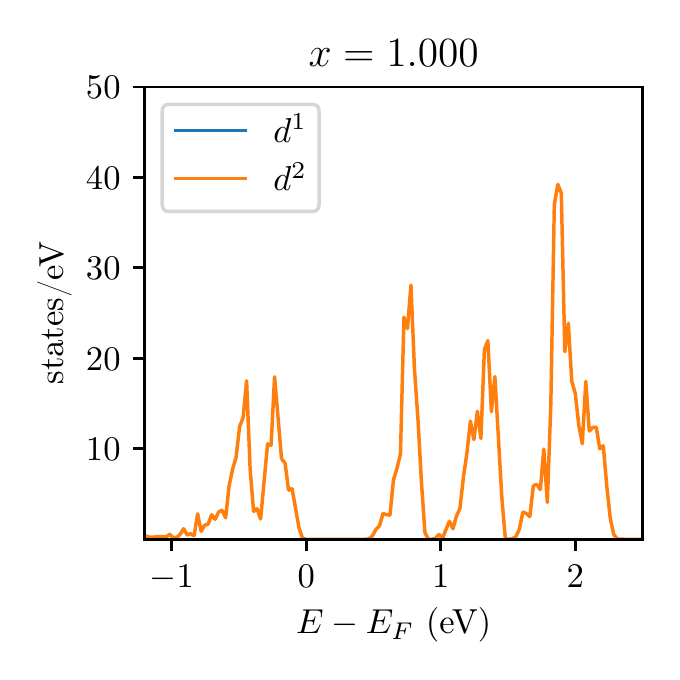}
\end{subfigure}
\caption{DOS projected onto the d-orbitals of Os $\mathrm{d^1}$ (blue) and $\mathrm{d^2}$ (orange) sites at different Ca concentration x.}
\label{figsm:dosx}
\end{figure}

\clearpage
\section{NMR anomalous peak fit results}
\label{secsm:fitresults}

We have fitted the NMR $1/T_1$ anomalous peak using the BPP-like model of Eq.~(\ref{eq:fitmodel}).
The results are reported in Tab.~\ref{smtab:fitresults}.

\begin{table}[ht]
\centering
\begin{tabular}{ccccc}
\hline
$x(\%)$ & $T_a$ (K)  & $E_a$ (meV) & $\tau_0$ (ps) & $\Delta^2 \times 10^9\ (s^{-2})$ \\ \hline
12.5    & $855\pm26$ & $74\pm2$    & $0.7\pm0.2$   & $28.0\pm0.8$                     \\
25.0    & $788\pm21$ & $68\pm2$    & $2.5\pm0.4$   & $27.5\pm0.6$                     \\
37.5    & $953\pm38$ & $82\pm3$    & $0.7\pm0.2$   & $97\pm3$                         \\
50.0    & $931\pm24$ & $80\pm2$    & $0.8\pm0.2$   & $129\pm3$                        \\
75.0    & $928\pm23$ & $80\pm2$    & $0.9\pm0.2$   & $143\pm3$                        \\
90.0    & $857\pm24$ & $77\pm2$    & $1.1\pm0.2$   & $97\pm2$                         \\ \hline
\end{tabular}
\caption{NMR anomalous peak fitting results}
\label{smtab:fitresults}
\end{table}

\clearpage
\section{Determination of the relaxation times $T_{1}$ and $T_{2}$}
\label{secsm:dynamicalcons}

The measure of $T_{1}$ is extracted by using the standard saturation recovery method with echo detection. In these conditions the nuclear spin transitions can be fully saturated but the detection reveals only the central transition.  In our sample $\mathrm{Ba_2Ca_{0.125}Na_{0.875}OsO_6}$, the NMR signal amplitude as a function of the repetition delay time t can be fitted to a stretched exponential function:
\begin{equation}
	M(t) = M_{0} \biggl( 1- exp \biggl(  - \biggl( \frac{t}{T_{1}}  \biggr)^{\beta_1}\biggr ) \biggr )
	\label{eqsm:T1expdecaybeta}
\end{equation}
where the common interpretation of $\beta$ is, in terms of the global relaxation, a system containing many independently relaxing species, resulting in the sum of different exponential decays.

The stretched behaviour is often observed in complex transition metal oxides. It typically reflects the presence of a non trivial distribution of relaxation rates due to local electronic inhomogeneities, which give rise to a site dependent magnetic or electric coupling.

A single exponential decay is clearly corresponding to $ \beta =1 $, while $\beta=0.5$ is typical of a fully disordered system,  \emph{i.e.} an intrinsic heterogeneity of phases, which can be described with a multi-exponential factor.

\begin{figure}[ht]
	\centering
	\includegraphics[width=0.4\linewidth,keepaspectratio]{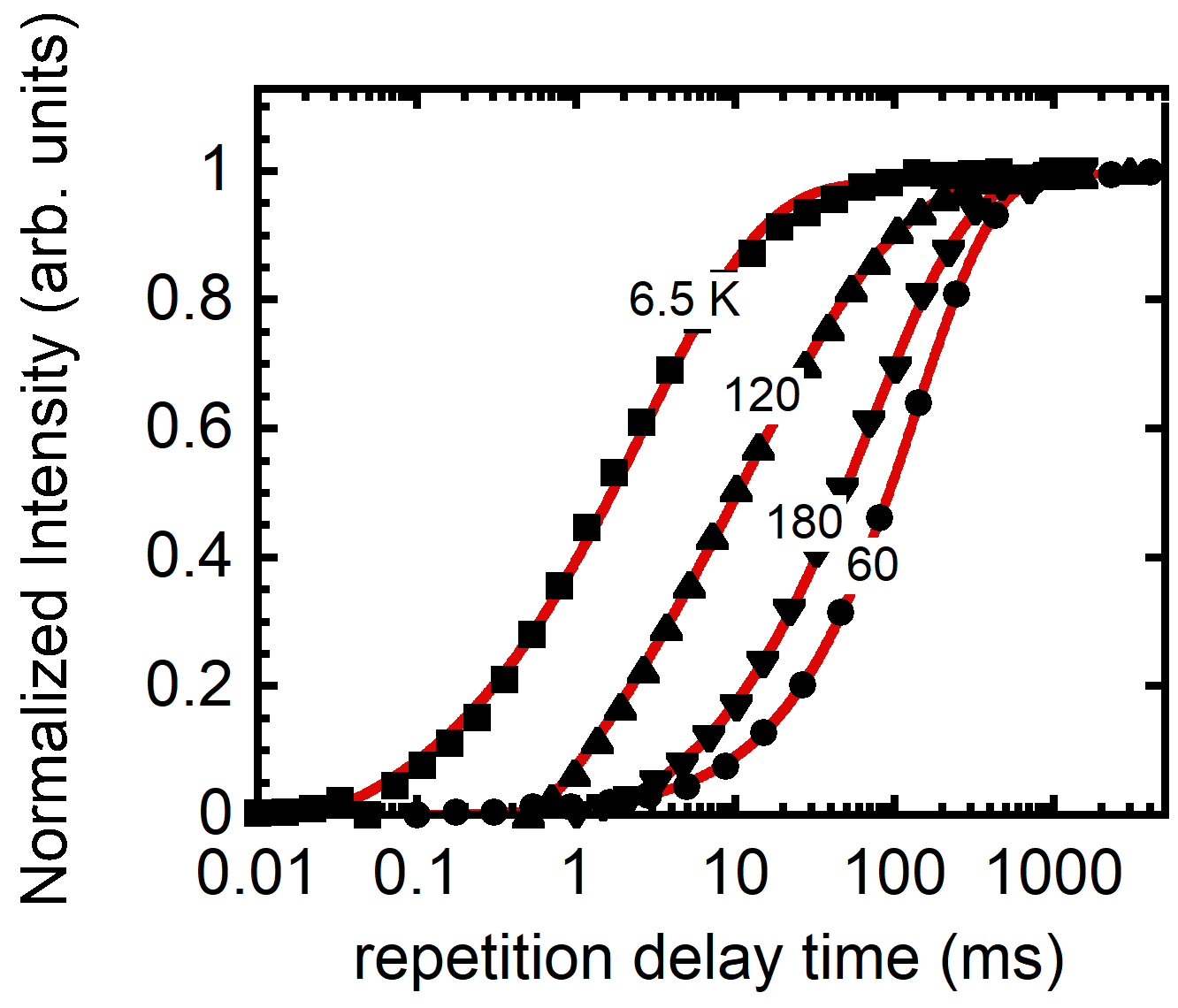}
	\caption{Plot of the normalized amplitude signal  as a function of the delay for the $T_{1}$ relaxation of the  $\mathrm{Ba_{2}Na_{0.125}Ca_{0.875}OsO_{6}}$ sample with the fit (solid line) to Eq.~(\ref{eqsm:T1expdecaybeta}).}
	\label{figsm:T1Stretched}
\end{figure}

Fig.~\ref{figsm:T1Stretched} shows the experimental behaviour with the fit line of Eq.~(\ref{eqsm:T1expdecaybeta}), for the normalized amplitude as a function of the delay time, for representative temperatures.

\begin{figure}[ht]
    \centering
    \includegraphics[width=0.4\linewidth]{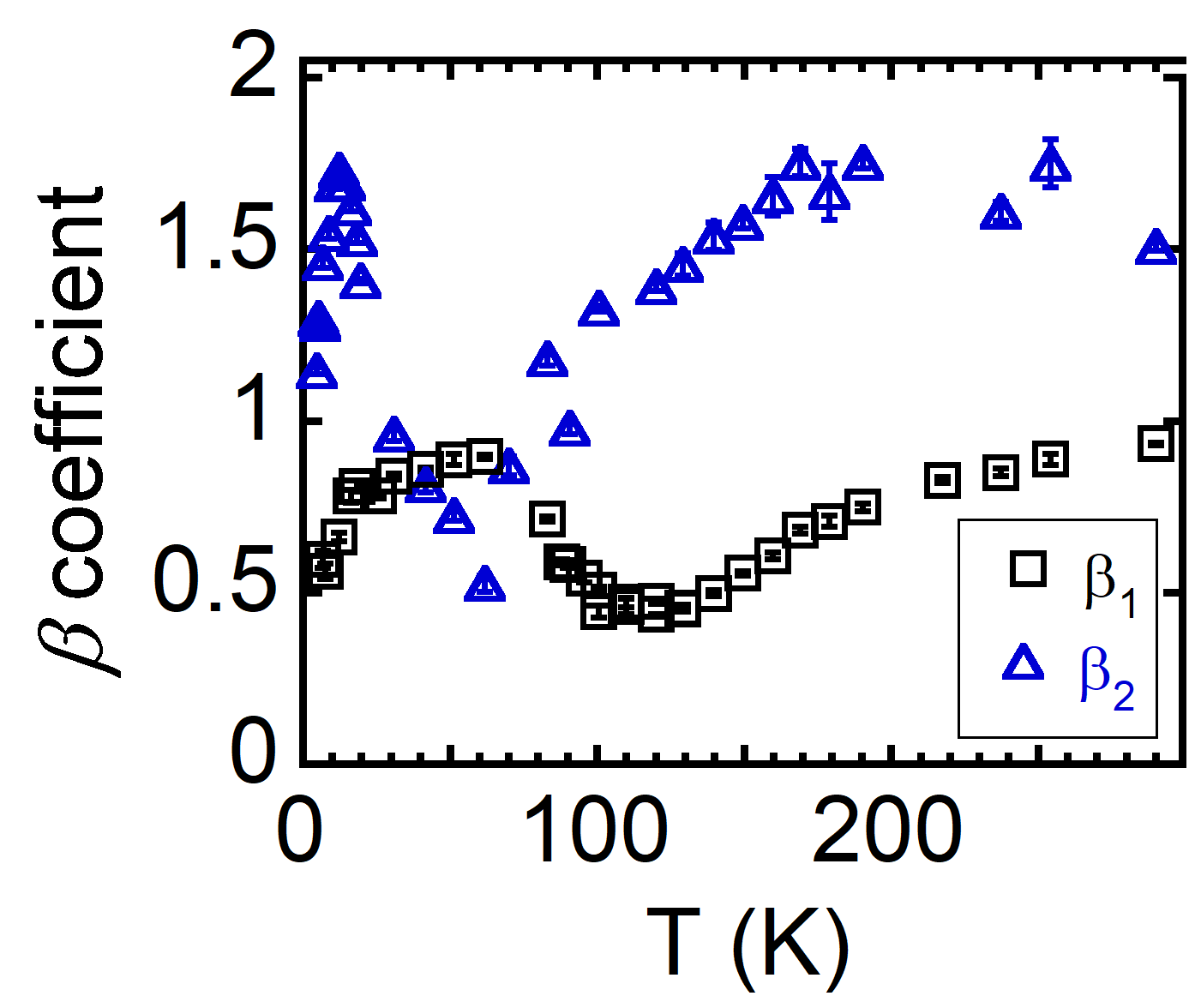}
    \caption{Temperature dependence of the $\beta_{i}$  exponent for $i=1,2$ of Eq.~(\ref{eqsm:T1expdecaybeta}) and Eq.~(\ref{eqsm:fitgenT2beta}), respectively.}
    \label{figsm:beta}
\end{figure}

The relaxation rate $1/T_1$ if displayed in the main text and the $\beta_1$ coefficient as a function of temperature in the whole range are reported in Fig.~\ref{figsm:beta} (squares). 
Notice that the stretching coefficient is reduced to $\beta=0.5$ at the relaxation rate peaks, revealing an electronic inhomogeneity \emph{i.e.} a $T_1$ distribution, but it approaches an exponential relaxation with $\beta=1$ elsewhere. 
The latter agrees with a quadrupolar relaxation mechanism as shown in Fig.~\ref{fig:doping} and Sec.~\ref{secsm:fitresults}.   

The measurements for the determination of the $ T_{2} $ transverse relaxation time have been performed by using a modified $ \pi $/2- $ \pi $/2 Hanh echo sequence).

The data has been fitted to a stretched exponential fit function, analogous to the one used for the $T_{1}$ longitudinal relaxation case in Eq.~(\ref{eqsm:T1expdecaybeta}), in the form:
\begin{equation}
	M(2\tau) = M_{0} \; exp \biggl( - \biggl( \frac{2\tau}{T_{2}}\biggr)^{\beta_2} \biggr),
	\label{eqsm:fitgenT2beta}
\end{equation}
with $\beta$ ranging from 0.5 to 2 reflecting a more dynamical disordered or more Gaussian character, respectively.

\begin{figure}[ht]
	\centering
	\includegraphics[width=0.4\linewidth, keepaspectratio]{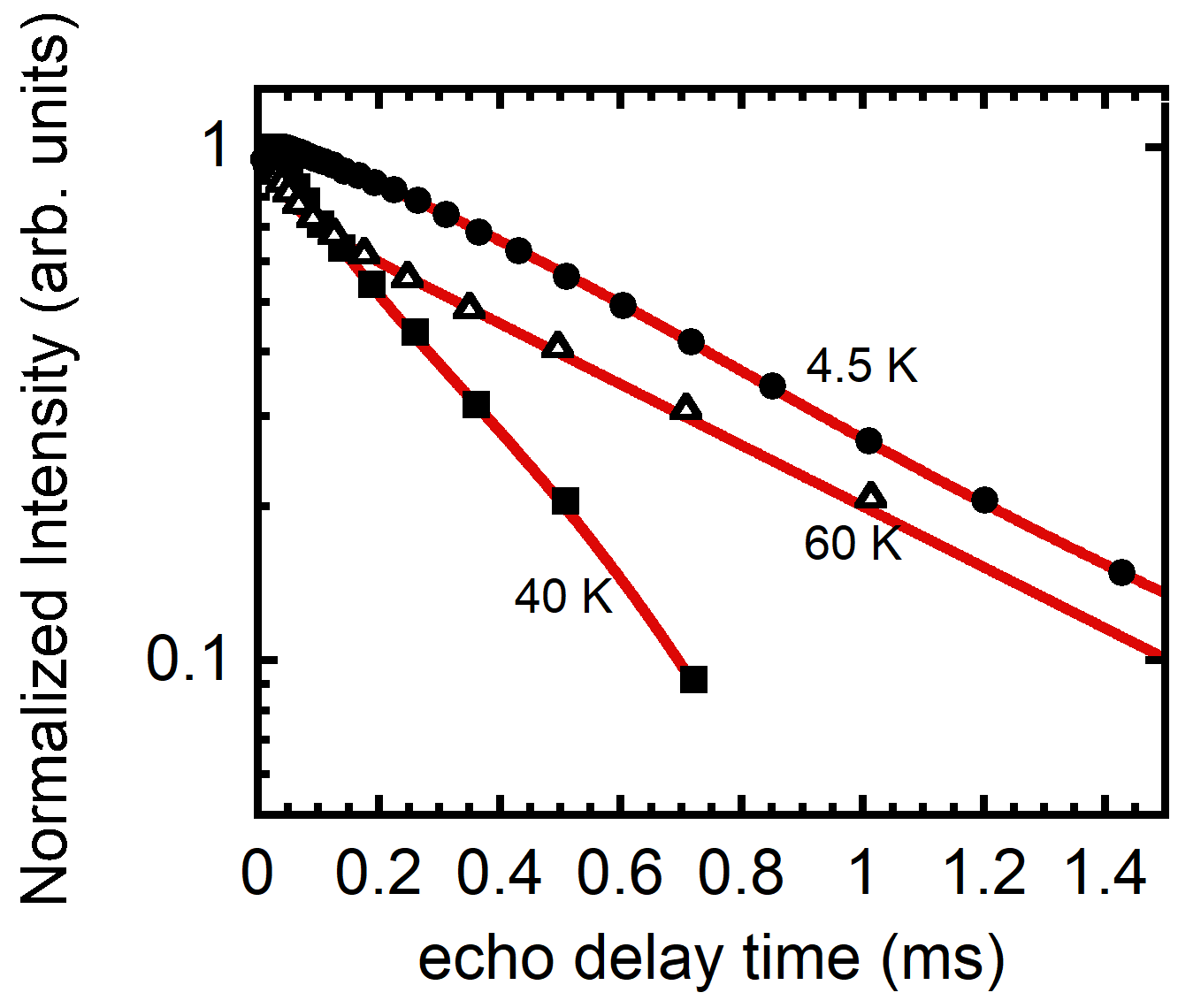}
	\caption{Plot of the normalized amplitude signal as a function of the  echo delay time for the $T_{2}$ relaxation of the $\mathrm{Ba_{2}Na_{0.125}Ca_{0.875}OsO_{6}}$ sample, for selected temperatures with the fit (solid line) to Eq.~(\ref{eqsm:fitgenT2beta}).}
	\label{figsm:T2Stretched}
\end{figure}

In Fig.~\ref{figsm:T2Stretched} is represented the fit function, expressed in the Eq.~(\ref{eqsm:fitgenT2beta}), for the normalized amplitude as a function of the echo delay time, for representative temperatures.

The relaxation rate $1/T_2$ if displayed in the main text and the $\beta_2$ coefficient as a function of temperature in the whole range are reported in Fig.~\ref{figsm:beta} (triangles).

The temperature dependence of the $\beta$ parameter, both from $T_1$ and $T_2$, reflects the system dynamics. We can observe that local minima of $\beta$ coefficients correspond to the peaks maxima of $1/T_{1}$ and $1/T_{2}$. This is expected in systems with a high electronic inhomogeneity which gives rise to a unresolved large distribution of correlation times.

\clearpage
\section{Muon Spin Rotation}
\label{secsm:musr}

Muon spin relaxation measurements have been performed at the GPS instrument at the Paul Scherrer Institute (Switzerland) in both zero field (ZF), $\mu_0$H=0, and longitudinal field (LF) conditions, where the latter uses an external field $\mu_0$H parallel to the initial muon spin polarization.

After the implantation into powder samples of a beam of completely polarized positively charged muons (spin $S_{\mu}$=1/2), we study the time evolution of muon spin asymmetry which provides information on the spatial distribution and dynamical fluctuations of the magnetic environment.

The $\mu SR$ is a magnetic probe and allows to investigate local magnetic moments values, but is not sensitive to charge variations, because muons have no electric quadrupole moment and are typically insensitive to the static or dynamical effects of the EFG.

\begin{figure}[ht]
    \centering
    \includegraphics[width=0.5\linewidth]{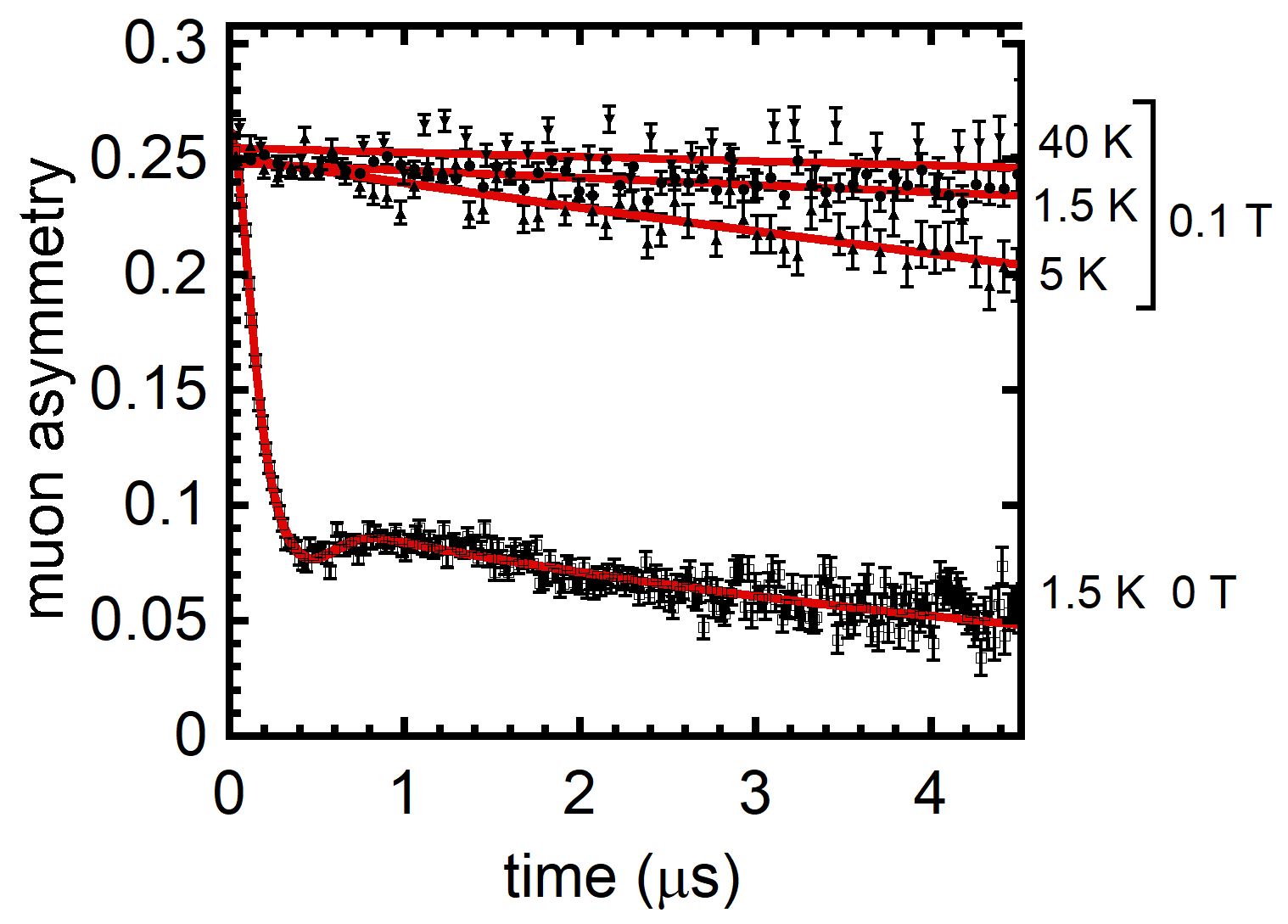}
    \caption{Representative time dependence of the muon decay asymmetry with fit curves (lines) to Eq.~(\ref{eqsm:asymm}): open squares are for external zero field condition at 1.5~K and solid symbols are for longitudinal applied field of $\mu_0H=0.1$~T at selected temperatures (circle 1.5~K, up triangles 5 K, down triangles 40~K). Solid line are the best fit to Eq.~(\ref{eqsm:asymm}) and Eq.~(\ref{eqsm:LFasymm}) for ZF and LF, respectively}
    \label{figsm:muonasy}
\end{figure}

The implanted muons decay with a characteristic lifetime of 2.2~$\mu$s, emitting a positron preferentially along the direction of the muon spin. The positrons are detected and counted by a forward ($N_F (t)$) and backward detector ($N_B (t)$) as a function of time. The asymmetry function A(t) is given by
\begin{equation}
A(t)=\frac{N_B (t) - \alpha N_F (t)}{N_B (t)+\alpha N_F (t)},
\end{equation}
where $\alpha$ is a parameter determined experimentally from the geometry and efficiency of the $\mu $SR detectors. $A(t)$ is proportional to the muon spin polarization, and thus reveals information about the local magnetic field sensed by the muons.
Examples of the muon asymmetry behavior are displayed in Fig.~\ref{figsm:muonasy}. The zero field spectra are those previously reported in ref.~\cite{garcia2022} and accordingly in the magnetic phase each individual spectra was fitted to a sum of precessing and relaxing asymmetries given by
\begin{eqnarray}
A(t)&= \left [A_1 e^{-\frac{\sigma_1^2 t^2}{2}}cos(2\pi \nu_1 t) + A_2 e^{-\frac{\sigma_2^2 t^2}{2}} \right ] + A_\ell e^{-{\lambda_\mu t}}
\label{eqsm:asymm}
\end{eqnarray}
The terms inside the brackets reflect the perpendicular component of the internal local field probed by the spin-polarized muons, the first term corresponds to the damped oscillatory muon precession about the local internal fields at frequencies $\nu_i$, while the second reflects a more incoherent precession with a local field distribution given by $\sigma$, for a total of 2 different muon sites (accidentally the most general case accounts up to three inequivalent muon sites for other compositions of the same series~\cite{garcia2022}). The term outside the brackets reflects the longitudinal component characterized by the muon spin-lattice relaxation rate $\lambda_\mu\equiv1/T_1^\mu$. 

LF-$\mu$SR measurements have been performed as a function of temperature in order to apply a static field for two field of 10 and 100~mT being the latter much greater than the internal static field detected in the ordered phase $\sim 10$~mT as reported in ref.\cite{garcia2022}. In this condition the longitudinal amplitude (the tail) is fully recovered to the maximum amplitude already for few tens of mT also at the base temperature indicating a static character of the magnetic state.  All the LF muon asymmetry data can be simply fitted to

\begin{equation}
A_\ell e^{-{\lambda_\mu t}} 
\label{eqsm:LFasymm}
\end{equation}

Both the ZF and LF longitudinal rates are reported in Fig.~\ref{fig:hopping}b as a function of temperature. They clearly show a relaxation peak due to critical fluctuations when approaching the magnetic transition at $T_N\simeq 7$~K. No evidence of extra anomalous relaxation peak is detected at any temperature up to 300~K.   

\clearpage
\section{Modelling of spin-lattice relaxation rate}
\label{secsm:model}

The spherical coordinate of the EFG appearing in Eq.~(\ref{eq:quadH}) are given in cartesian coordinates by~\cite{mehring_principles_1983}
\begin{subequations}
    \begin{align}
        &V_0 = \frac{1}{\sqrt{6}} \left[ 3V_{zz} - \left( V_{xx} + V_{yy} + V_{zz}\right) \right] \label{eqsm:pot1} \\ 
        &V_{\pm 1} = \mp\frac{1}{2}\left[V_{xz} + V_{zx} \pm i\left( V_{yz} + V_{zy}\right) \right] \label{eqsm:pot2} \\
        &V_{\pm 2} = \frac{1}{2}\left[V_{xx} - V_{yy} \pm i\left( V_{xy} + V_{yx}\right) \right] \label{eqsm:pot3}
    \end{align}
    \label{eqsm:potsph}
\end{subequations}
where the lower indices indicate derivatives with respect to the corresponding cartesian coordinate.
To estimate the EFG we used a point-charge model of the NaO$_6$ octahedron with a coulombic potential given by
\begin{equation}
    V(\bm{r}) = q_{ox}\sum_{i=1}^6 \frac{1}{|\bm{r} - \bm{R}_i|}
    \label{eqsm:potential}
\end{equation}
where $q_{ox}$ is the formal charge of the oxagen ions and $\bm{R}_i$ is the position of the $i$-th oxygen ion with respect to the nucleus. 
For the oxygen labelling convention see Fig.~\ref{figsm:qs}.

By combining Eq.~(\ref{eqsm:potsph}) and Eq.~(\ref{eqsm:potential}) we can calculate the matrices $M^{(ij)}_{\alpha\beta}$ used in the derivation of the spin-lattice relaxation rate $1/T_1$ in the Methods section.

\begin{equation}
    M^{(11)} = M^{(44)} = \frac{1}{R^8_0}
    \begin{pmatrix}
        3 & 0 & 0 \\ 0 & 3 & 0 \\ 0 & 0 & 9
    \end{pmatrix} \qquad
    M^{(22)} = M^{(55)} =  \frac{1}{R^8_0}
    \begin{pmatrix}
        9 & 0 & 0 \\ 0 & 3 & 0 \\ 0 & 0 & 3
    \end{pmatrix} \qquad
    M^{(33)} = M^{(66)} = \frac{1}{R^8_0}
    \begin{pmatrix}
        3 & 0 & 0 \\ 0 & 9 & 0 \\ 0 & 0 & 3
    \end{pmatrix} \qquad
\end{equation}

\begin{equation}
 M^{(14)} = \frac{1}{R^8_0}
    \begin{pmatrix}
        -3 & 0 & 0 \\ 0 & -3 & 0 \\ 0 & 0 & -9
    \end{pmatrix} \qquad
 M^{(25)} = \frac{1}{R^8_0}
    \begin{pmatrix}
        -9 & 0 & 0 \\ 0 & -3 & 0 \\ 0 & 0 & -3
    \end{pmatrix} \qquad
 M^{(36)} = \frac{1}{R^8_0}
    \begin{pmatrix}
        -3 & 0 & 0 \\ 0 & -9 & 0 \\ 0 & 0 & -3
    \end{pmatrix}
\end{equation}

\begin{equation}
 M^{(12)} = -M^{(15)} = -M^{(42)} = M^{(45)} = \frac{1}{R^8_0}
    \begin{pmatrix}
        0 & 0 & 3 \\ 0 & 0 & 0 \\ -9/2 & 0 & 0
    \end{pmatrix}
\end{equation}

\begin{equation}
    M^{(13)} = -M^{(16)} = -M^{(43)} = M^{(46)} = 
    \frac{1}{R^8_0}  
    \begin{pmatrix}
        0 & 0 & 0 \\ 0 & 0 & -3 \\ 0 & -9/2 & 0
    \end{pmatrix}
\end{equation}

\begin{equation}
 M^{(23)} = -M^{(26)} = -M^{(53)} = M^{(56)} = 
 \frac{1}{R^8_0}
    \begin{pmatrix}
        0 & -9/2 & 0 \\ 3 & 0 & 0 \\ 0 & 0 & 0
    \end{pmatrix} \qquad
\end{equation}

\clearpage
\printbibliography
\end{refsection}

\end{document}